\documentclass[]{spie}  

\usepackage[usenames,dvipsnames]{color}
\usepackage{ulem}
 
\usepackage{amsmath,amsfonts,amssymb}
\usepackage{graphicx} 
\graphicspath{{images/}} 
\usepackage[lofdepth,lotdepth,subrefformat=parens,labelformat=parens]{subcaption}
\usepackage[colorlinks=true, allcolors=blue]{hyperref}

\usepackage[T1]{fontenc}
\usepackage[utf8]{inputenc}
\usepackage{babel}
\usepackage{CJKutf8}

\usepackage{siunitx}
\newcommand{\SIadj}[2]{\SI[number-unit-product={\text{-}}]{#1}{#2}}
\usepackage[acronym]{glossaries}
\usepackage{glossaries-extra}

\usepackage{multicol}
\usepackage{boldline}

\usepackage{xcolor}

\setabbreviationstyle[acronym]{long-short}

\glssetcategoryattribute{acronym}{nohyperfirst}{true}

\newacronym{metal}{MetaL}{metamaterial-based lenslet}
\newacronym{cmb}{CMB}{cosmic microwave background}
\newacronym{fss}{FSS}{frequency selective surface}
\newacronym{srr}{SRR}{split-ring resonator}
\newacronym{nist}{NIST}{National Institute of Science \& Technology}
\newacronym{kid}{KID}{kinetic inductance detector}
\newacronym{cobe}{COBE}{COsmic Background Explorer}
\newacronym{wmap}{WMAP}{Wilkinson Microwave Anisotropy Probe}
\newacronym{gut}{GUT}{grand unified theory}
\newacronym{bao}{BAO}{baryonic acoustic oscillations}
\newacronym{lcdm}{$\Lambda$CDM}{$\Lambda$ cold dark matter}
\newacronym{des}{DES}{dark energy survey}
\newacronym{so}{SO}{Simons Observatory}
\newacronym{lat}{LAT}{large aperture telescope}
\newacronym{sat}{SAT}{small aperture telescope}
\newacronym{lbird}{LiteBIRD}{Lite satellite for the study of B-mode polarization and Inflation from cosmic background Radiation Detection}
\newacronym{lft}{LFT}{low frequency telescope}
\newacronym{mft}{MFT}{medium frequency telescope}
\newacronym{hft}{HFT}{high frequency telescope}
\newacronym{jaxa}{JAXA}{Japan Aerospace Exploration Agency}
\newacronym{lmt}{LMT}{Large Millimetre Telescope}
\newacronym{atran}{ATRAN}{Atmospheric TRANsmission}
\newacronym{hwp}{HWP}{half-wave plate}
\newacronym{fpu}{FPU}{focal plane unit}
\newacronym{snr}{SNR}{signal-to-noise ratio}
\newacronym{nep}{NEP}{noise equivalent power}
\newacronym{omt}{OMT}{ortho-mode transducer}
\newacronym{spt}{SPT}{South Pole Telescope}
\newacronym{act}{ACT}{atacama cosmology telescope}
\newacronym{class}{CLASS}{Cosmology Large Angular Scale Surveyor}
\newacronym{arc}{ARC}{anti-reflection coating}
\newacronym{blast-tng}{BLAST-TNG}{Balloon-borne Large Aperture Submillimeter Telescope – The Next Generation}
\newacronym{nika}{NIKA}{New IRAM KID Array}
\newacronym{scuba}{SCUBA}{Submillimetre Common-User Bolometer Array}
\newacronym{jcmt}{JCMT}{James Clerk Maxwell Telescope}
\newacronym{pixie}{PIXIE}{Primordial Inflation Explorer}
\newacronym{tes}{TES}{transition edge sensor}
\newacronym{mux}{MUX}{multiplexing}
\newacronym{spider}{SPIDER}{Suborbital Polarimeter for Inflation Dust and the Epoch of Reionisation}
\newacronym{bicep}{BICEP}{Background Imaging of Cosmic Extragalactic Polarization}
\newacronym{polarbear}{POLARBEAR}{Polarization of Background Radiation}
\newacronym{piper}{PIPER}{Primordial Inflation Polarization Explorer}
\newacronym{tod}{TOD}{time ordered data}
\newacronym{lgb}{LGB}{Laguerre-Gaussian beam}
\newacronym{oa}{OA}{optical axis}
\newacronym{fov}{FOV}{field of view}
\newacronym{fwhm}{FWHM}{full width half maximum}
\newacronym{camb}{CAMB}{Code for Anisotropies in the Microwave Background}
\newacronym{fft}{FFT}{fast Fourier transform}
\newacronym{ifft}{IFFT}{inverse fast Fourier transform}
\newacronym{rms}{RMS}{root mean square}
\newacronym{tl}{TL}{trasmission line}
\newacronym{pei}{PEI}{polyetherimide}
\newacronym{tem}{TEM}{transverse electro-magnetic}
\newacronym{lpf}{LPF}{low-pass filter}
\newacronym{hpf}{HPF}{high-pass filter}
\newacronym{bpf}{BPF}{band-pass filter}
\newacronym{hfss}{ANSYS-HFSS}{ANSYS High Frequency Structure Simulator}
\newacronym{mcmc}{MCMC}{Markov chain Monte Carlo}
\newacronym{drie}{DRIE}{deep reactive ion etch}
\newacronym{rie}{RIE}{reactive ion etch}
\newacronym{lpcvd}{LPCVD}{low pressure chemical vapour deposition}
\newacronym{tx}{Tx}{transmitter}
\newacronym{rx}{Rx}{receiver}
\newacronym{vna}{VNA}{vector network analyser}
\newacronym{ram}{RAM}{radiation absorbent material}
\newacronym{idc}{IDC}{inter-digital capacitor}
\newacronym{soi}{SOI}{silicon-on-oxide}
\newacronym{fem}{FEM}{finite element method}
\newacronym{fir}{FIR}{far infra-red}
\newacronym{sz}{SZ}{Sunyaev-Zeldovich}
\newacronym{net}{NET}{noise equivalent temperature}
\newacronym{cad}{CAD}{computer assisted design}
\newacronym{pp}{PP}{polypropylene plastic}
\newacronym{febecop}{FEBeCoP}{Fast Effective Beam Convolution in Pixel space}
\newacronym{dut}{DUT}{device under test}
\newacronym{if}{IF}{intermediate frequency}
\newacronym{bcs}{BCS}{Bardeen–Cooper–Schrieffer}
\newacronym{gr}{GR}{generation-recombination}
\newacronym{lekid}{LeKID}{lumped-element kinetic inductance detector}
\newacronym{rf}{RF}{radio frequency}
\newacronym{lna}{LNA}{low noise amplifier}
\newacronym{tls}{TLS}{two level system}
\newacronym{cpw}{CPW}{co-planar waveguide}
\newacronym{ms}{MS}{microstrip}
\newacronym{ads}{ADS}{Advanced Design System}
\newacronym{iq}{IQ}{in-phase/quadrature}
\newacronym{lo}{LO}{local oscillator}
\newacronym{psd}{PSD}{power spectral density}
\newacronym{ptc}{PTC}{pulse tube cryocooler}
\newacronym{ebpvd}{EBPVD}{electron-beam physical vapor deposition}
\newacronym{nmp}{NMP}{N-methyl-2-pyrrolidone}
\newacronym{ipa}{IPA}{isopropyl-alcohol}
\newacronym{icp}{ICP}{inductively coupled plasma}
\newacronym{bbn}{BBN}{Big Bang nucleosynthesis}
\newacronym{wimp}{WIMP}{weakly interacting massive particle}
\newacronym{ccat}{CCAT}{Cerro Chajnantor Atacama Telescope}
\newacronym{ot}{OT}{optics tube}
\newacronym{dr}{DR}{dilution refrigeration}
\newacronym{latr}{LATR}{large aperture telescope receiver}
\newacronym{fea}{FEA}{finite element analysis}
\newacronym{dof}{DOF}{degrees of freedom}
\newacronym{mli}{MLI}{multi-layered insulation}
\newacronym{fyst}{FYST}{Fred Young Submillimeter Telescope}

\title{Thermal and mechanical study of a parametrised cryostat model for optical characterisation of upcoming CMB experiments}

\author[a]{Thomas J.L.J. Gascard}
\author[b]{Yi Wang (\begin{CJK*}{UTF8}{gbsn}王逸)\end{CJK*}}
\author[a]{Jon E. Gudmundsson}
\author[b,c]{Eve~M.~Vavagiakis}
\author[c]{Cody~J.~Duell}
\author[c]{Zachary~B.~Huber}
\author[c]{Lawrence~T.~Lin}
\author[c,d]{Michael~D.~Niemack}
\author[d]{Rodrigo~G.~Freundt}
\affil[a]{University of Iceland, School of Engineering and Natural Sciences, 107 Reykjavík, Iceland}
\affil[b]{Duke University, Department of Physics, Durham, NC 27708, United States of America}
\affil[c]{Cornell University, Department of Physics, Ithaca, NY, 14853, United States of America}
\affil[d]{Cornell University, Department of Astronomy, Ithaca, NY, 14853, United States of America}

\authorinfo{Further author information: (Send correspondence to Thomas Gascard)\\Thomas Gascard: E-mail: thomasgascard@hi.is}

\begin{document} 
\maketitle

\begin{abstract}
Current and future experiments observing the cosmic microwave background require a detailed understanding of optical performance at cryogenic temperatures. Pre-deployment analysis of optics can be performed in custom-engineered cryogenic test beds, such as Mod-Cam, a first light camera for the CCAT project. This work presents studies of the mechanical and thermal performance of CryoSim, a model of a generic cylindrical \SIadj{4}{K} cryostat cooled with a commercial pulse tube cryocooler that can be used to characterise optical components and full reimaging optical systems. CryoSim is extensively parametrised, allowing the joint analysis and optimisation of mechanical and thermal performance via finite element methods. Results from this model are validated against measured cooldown data of the Mod-Cam cryostat. Due to the extensive parametrisation of the model, significant modifications of the cryostat geometry may be implemented to be representative of any system the scientific community may desire, and validation of thermal and mechanical performance can be carried out rapidly.
\end{abstract}



\section{INTRODUCTION}
\label{sec:intro}  

\subsection{Context}
\label{subsec:intro_context}

Observations of the \gls{cmb} have had tremendous impact on the development on modern cosmology.\cite{Planck18-6} Current and future generation instruments are designed to engender deep maps of the polarisation of the \gls{cmb} across the entire sky with high-fidelity\cite{ade_simons_2019, hazumi_litebird_2019}. Continued measurements at millimetre wavelengths can provide a unique insight into fundamental physics and the early universe while constraining extensions to $\Lambda CDM$ cosmology and tracing galaxy evolution over cosmic time \cite{abazajian_cmb-s4_2016, CCAT_Prime_Collaboration_2022}.


Experiments mapping the \gls{cmb} rely on either refracting or reflecting telescopes. Reflecting telescopes typically use mirrors at ambient temperatures that feed into cryogenically-cooled reimaging optics\cite{thornton_atacama_2016}, while refracting telescopes house the majority of their optical elements inside an actively cooled \gls{ot}, standardised modules composed of various optical elements such as filters, lenses, and a detector focal plane.\cite{ade_bicep2_2015} Regardless of the overall design, a detailed understanding of the electromagnetic properties of cryogenically cooled optical elements is required for upcoming experiments to achieve their science goals\cite{chesmore_simons_2022}. Many cryogenic tests beds dedicated to such calibration of future \gls{cmb} experiments rely on one or several \glspl{ptc} that cool a cylindrical cryostat housing a single \gls{ot}\cite{galitzki2024simons}. An example of such a system is the Mod-Cam test bed built for the CCAT project\cite{vavagiakis_ccatprime_2022}. In recent years, large systems have been designed to accommodate multiple optics tubes \cite{thornton_atacama_2016,zhu_simons_2021}. The design phase for these cryostats often involves complex optimization problems with figures of merit that can sometimes be difficult to quantify. This type of analysis can involve considerations regarding thermally insulating support structures, overall system volume, cryocooler cooling capacities, number of temperature stages, temperature gradients, thermal transients etc.\

In an attempt to help speed up initial steps of this process, we have developed CryoSim, an extensively parameterised mechanical and thermal model of a test cryostat, implemented in Solidworks and COMSOL, which can serve to inform the design of an optical test bed for future \gls{cmb} experiments. An overview of the model considered and the iterative procedure carried out to establish a reliable design is discussed in Section~\ref{subsec:overview_model}. The full process is summarised in Section~\ref{sec:prelim_gencryo}. It starts with a preliminary mechanical study of a simplified model of the cryostat, further discussed in Section~\ref{subsec:mecha_gencryo}. Once the reduced design converges on acceptable stresses and natural vibrations, a thermal analysis is carried out to validate that temperature gradients are within a satisfactory range, as will be shown in Section~\ref{subsec:thermal_gencryo}. If modifications are necessary at this stage, the procedure restarts and the cycle is repeated until adequate performance is reached. To showcase the parametrisation capabilities of the CryoSim model, a study on the reduction of thermally isolating mechanical supports used to link cryogenic stages is presented in Section~\ref{sec:case1_g10}. This model will be used to advance the design of a test cryostat that will be built and located at the University of Iceland. Preparing for the latter motivated the development of CryoSim, but it is also hoped that this work can be useful to the wider \gls{cmb} community. The COMSOL and Solidworks model, material database, and codes use throughout this work are publicly available on the GitHub \href{https://github.com/Skuggsja-Lab/skuggsja-cryobeam/}{repository} of the University of Iceland's laboratory.

\subsection{Model and procedure overview}
\label{subsec:overview_model}
The CryoSim model is based on Mod-Cam, a single \gls{ot} test bed and first light instrument for the CCAT project's \gls{fyst}, which is currently in testing at Cornell University \cite{vavagiakis_ccatprime_2022}. The preliminary temperature gradients, measured dark with the cryostat cooled down to \SI{4}{K} without the \gls{dr} extension or the \gls{ot}, will be compared with the results of the thermal model in Section~\ref{subsec:model_rely} as a validation step.

\begin{figure}[h!]
	\centering
	\begin{subfigure}[b]{0.39\textwidth}
		\centering
		\includegraphics[width=\textwidth]{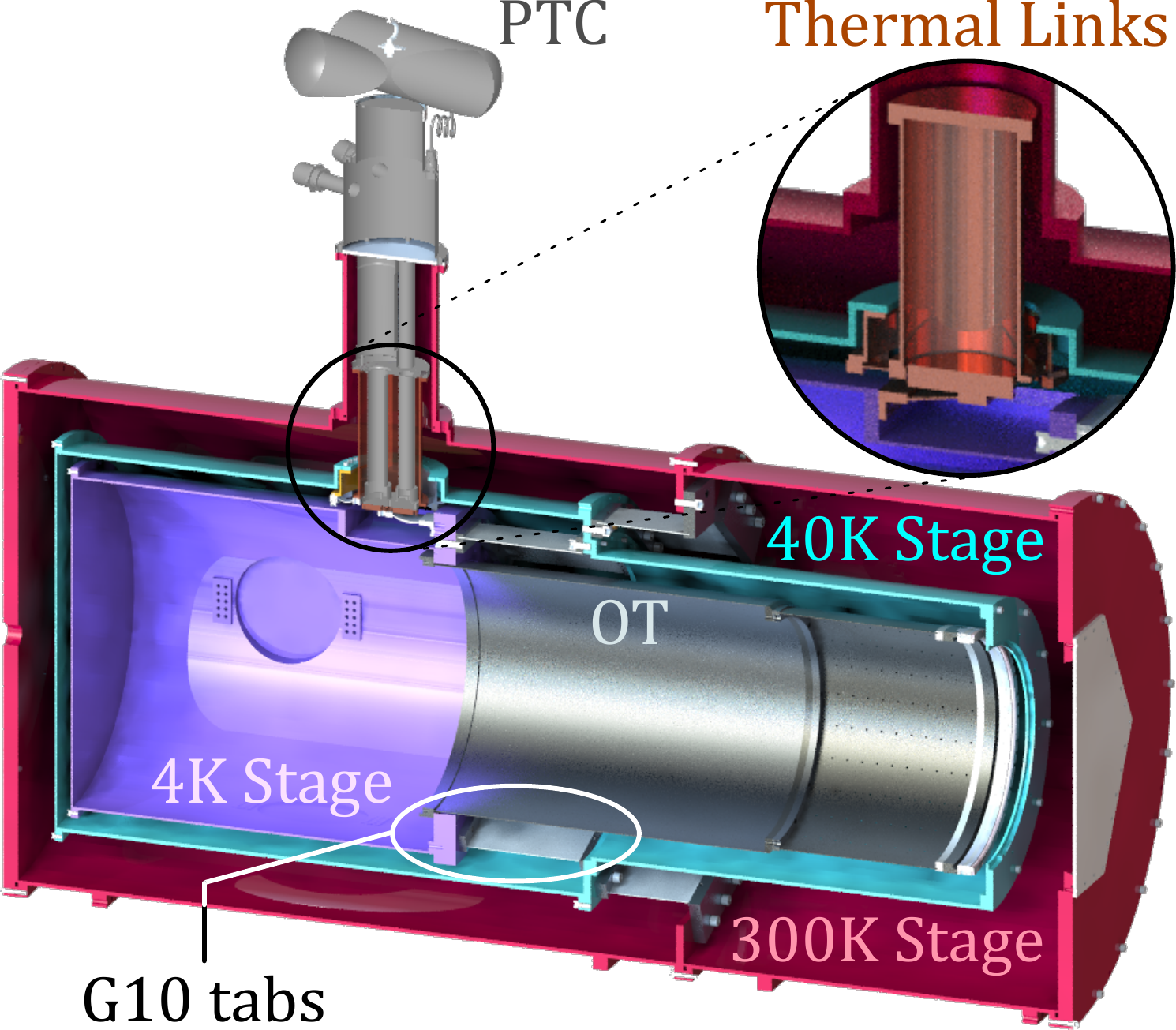}
		\caption{The Mod-Cam cryostat}
		\label{subfig:cryostats_render_1}
	\end{subfigure}
	\hfill
	\begin{subfigure}[b]{0.47\textwidth}
		\centering
		\includegraphics[width=\textwidth]{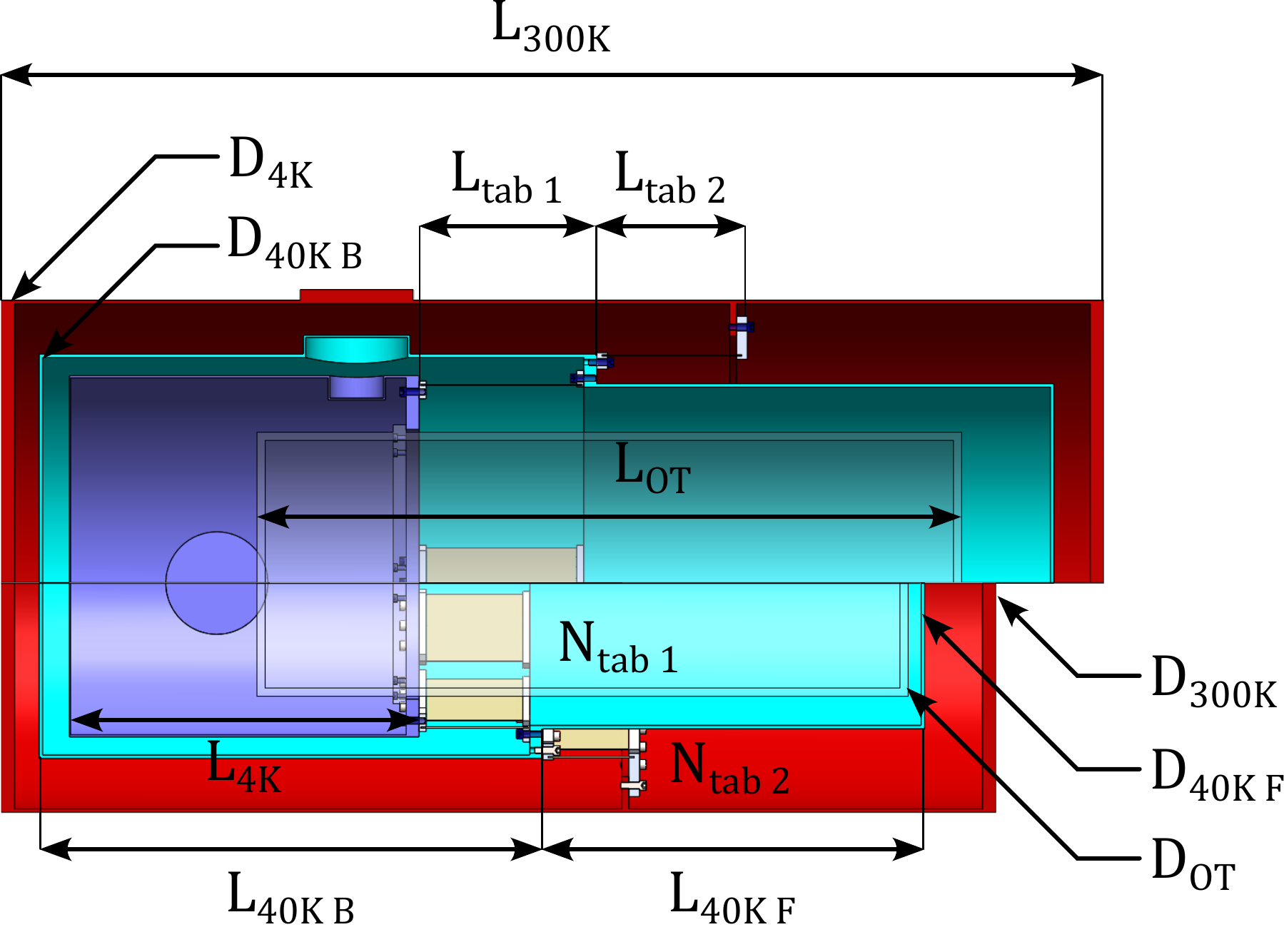}
		\caption{The CryoSim model}
		\label{subfig:cryostats_render_2}
	\end{subfigure}
    \caption{\protect\subref{subfig:cryostats_render_1} Rendering of the Mod-Cam cryostat. Cooling power is provided by the \gls{ptc} installed atop the cryostat and connected with thermal links at the \SIadj{40}{K} and \SIadj{4}{K} stages. Each shell is connected with the preceding one through a set of G10 fibreglass tabs. \protect\subref{subfig:cryostats_render_2} A split view of the the CryoSim model. The bottom half is set up in a configuration based on Mod-Cam. The top half illustrates a different case for which a selection of parameters ($D_{\text{OT}}$, $L_{\SI{4}{K}}$, $N_{\text{tab 1}}$, etc.) are changed.}
    \label{fig:cryostats_render}
\end{figure}

Figure~\ref{fig:cryostats_render} compares the Mod-Cam CAD rendering to the nominal model implemented in CryoSim. It consists of a \SIadj{300}{K}, \SIadj{40}{K} and \SIadj{4}{K} stage, surrounding the \gls{ot} and connected via G10 fibreglass tabs \cite{zhu_simons_2021, galitzki_simons_2024}. The \gls{ot} is initally incorporated for the mechanical analysis but is then removed for the thermal study. The cooling power is provided by a \href{https://bluefors.com/products/pulse-tube-cryocoolers/pt420-pulse-tube-cryocooler/}{PT420} from Bluefors-Cryomech attached on top the cryostat. On Mod-Cam, the \SIadj{40}{K} stage is shielded with \gls{mli} to absorb any stray coming through the window at the front of the \SIadj{300}{K} shell and mitigate radiative heat transfer from the latter. Each cooling stage of the \gls{ptc} is thermally linked via dedicated copper connections. CryoSim is fully parametrised to enable rescaling and adjustments to accommodate for a variety of \gls{ot} configurations, or help with any desired modifications. While being representative of a real cryostat, simplified geometry is essential in generating a sensible mesh for the \gls{fem} analysis being conducted. The mechanical stress and natural frequencies studies are conducted in Solidworks (Education Edition, 2022 - 2023, SP5.0). The thermal analysis is done in COMSOL Multiphysics (Version 6.1, Build 357, with Livelink for Solidworks and Heat Transfer modules).

The CryoSim model, shown in Fig.~\ref{fig:cryostats_render}, is fully parametrised. For an optics tube of a given diameter, length and weight, the geometry of each stage, their mechanical supports and the thermal connections can be tuned at will. The Solidworks and COMSOL multiphysics models are linked, allowing for a joint mechanical and thermal analysis. The iterative optimisation process is summarised as follows: 
\begin{itemize}
    \item Mechanical refinement and subsequent validation of the geometry is done in Solidworks by running static stress and modal analysis. The model is parametrised to allow for modification of a great number of properties, including: the number, position and dimensions of the G10 tabs, the \gls{ot} length and diameter, the clearance between each successive stage, the position of the \gls{ptc}, etc.
    \item Once the mechanical concept is validated, a stationary thermal analysis is run in COMSOL Multiphysics. This provides a set of cryogenic figures of merit over the  full system, such as the temperature gradients, thermal stage equilibrium temperatures, etc.
    \item The cryogenic figures of merit are used to guide design modifications, implemented in Solidworks, towards improving performance. 
\end{itemize}
These three steps are repeated iteratively until convergence on a design that meets all thermal and mechanical targets. In the following section, the first and second step of the design refinement procedure will be conducted on the CryoSim model as a preliminary design analysis, which will further demonstrate the capabilities of such joint approach.

\section{PRELIMINARY JOINT THERMAL AND MECHANICAL ANALYSIS}
\label{sec:prelim_gencryo}

\subsection{Mechanical analysis}
\label{subsec:mecha_gencryo}

\subsubsection{Mechanical model and simulation setup}
\label{subsubsec:mecha_setup}
\begin{figure}[h!]
    \centering
    \includegraphics[width=\textwidth, keepaspectratio]{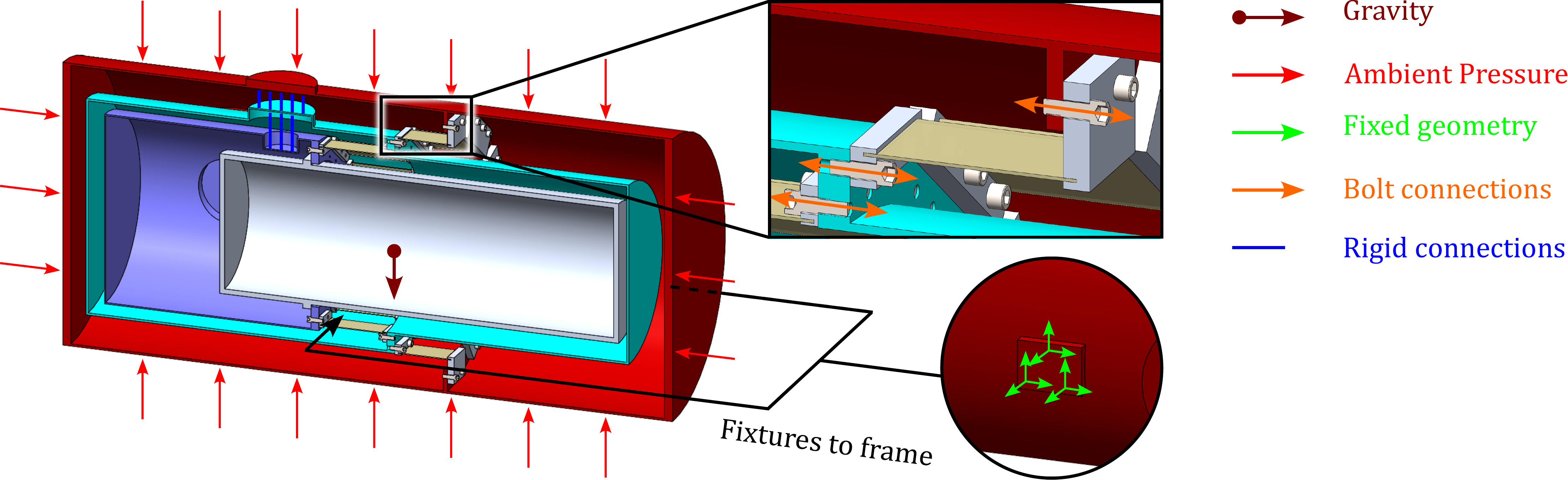}
    \caption{A cut view of the CryoSim model, displaying the stress and acoustic simulation setup, underlining connections, contacts and loads applied to the cryostat.}
    \label{fig:meca_sim_setup}
\end{figure}

The CryoSim mechanical model currently encompasses all 3 stages with the thermal links to the \gls{ptc}, the G10 tabs and a dummy geometry with the same form, center of mass and weight as a typical \gls{ot}. The analytical setup, depicted in Figure~\ref{fig:meca_sim_setup}, consists of defining the external loads applied to the assembly. This includes the fixtures connecting different thermal stages, namely the G10 tabs and bolts, and fixtures external to the vacuum shell, bolted to a fiducial mounting frame from a protrusion on each side, modelled as a fixed geometry. The pulse tube was removed for this preliminary analysis, and replaced with rigid connections. The definition of the constraints and load types as specified in Solidworks can be found in the software \href{https://help.solidworks.com/2024/english/SolidWorks/SWHelp_List.html}{online help}.

In Solidworks, components are said to be in contact if they are physically touching and may slide against one another under load, without penetration. Internal bolts are incorporated automatically as a distribution of mesh nodes at the physical contact surface of the bolt's head, coupled to a reference central node at the shank. The sum of the forces at the coupling nodes is equivalent to the total pre-load at the reference node, leading to accurate stress and displacement fields at the connection surfaces. Perfectly bonded contacts, for which sliding is unauthorised, were defined for the part of the G10 tabs connected with their aluminium feet, a model that does not incorporate the non-linear characteristics of a glued assembly. Future update will include a non-linear static analysis of the CryoSim model, including an epoxy glue layer between the G10 tabs and their feet that will be modelled after measurements taken in Cornell University. The rest of the assembly is setup in normal contact at a global level, with soft spring stabilisation to help with the convergence.

The external loads considered are the ambient pressure resulting from pumping the cryostat, set at \SI{1}{atm}, and the gravity applied downwards.
Solidworks has pre-defined materials properties for a variety of aluminum alloys and copper. The G10 characteristics are compiled from various supplier datasheets. Table~\ref{tab:meca_matprops} summarises the relevant mechanical properties of the various material constitutive of the design. For the current pre-analysis, the meshes defined for the static loading and the acoustic analysis have the properties summarised in Table~\ref{tab:acou_meshprops} of Appendix~\ref{apxsubsec:meca_meshes}.
\begin{table}[h!]
	\centering
        \caption{Mechanical properties of Al 6063 and Al 6061 alloys, copper and G10. $\rho$ is the density, $\alpha_T$ is the coefficient of linear thermal expansion, $E$ is the elastic modulus, $\nu$ is the Poisson ratio, $\sigma_y$ is the yield strength and $\sigma_u$ is the tensile strength.}
	\resizebox{0.975\textwidth}{!}{
		\begin{tabular}{|l|c|c|c|c|c|c|c|}
			\hlineB{2} \hline
                Material & Elements & $\rho [\si{\kilogram\per\cubic\metre}]$ & $\alpha_T [\si{\per\kelvin}]$ & $E [\si{\mega\pascal}]$ & $\nu$ [\%] & $\sigma_y [\si{\mega\pascal}]$ & $\sigma_u [\si{\mega\pascal}]$ \\
                \hlineB{2} \hline
                Al 6063 (T6) & \SIadj{40}{K} Stage \& Tab feet & 2700 & \num{2.3e-5} & \num{6.9e+4} & 33 & 215 & 240 \\
                Al 6061 (O) & \SIadj{300}{K} \& \SIadj{4}{K} Stages & 2700 & \num{2.4e-5} & \num{6.9e+4} & 33 & 62 & 125 \\
                Cu & Thermal links & 8900 & \num{2.4e-5} & \num{1.1e+5} & 37 & 259 & 394 \\
                G10 & Tabs & 2440 & \num{0.9e-5} & \num{1.9e+4} & 12 & 276 (transverse) & 310 (transverse) \\
			\hlineB{2} \hline
		\end{tabular}
	}
\label{tab:meca_matprops}
\end{table}

\subsubsection{Stress and acoustic analysis}
\label{subsubsec:mecha_results}
\begin{figure}[h!]
	\centering
	\begin{subfigure}[b]{0.48\textwidth}
		\centering
		\includegraphics[width=\textwidth]{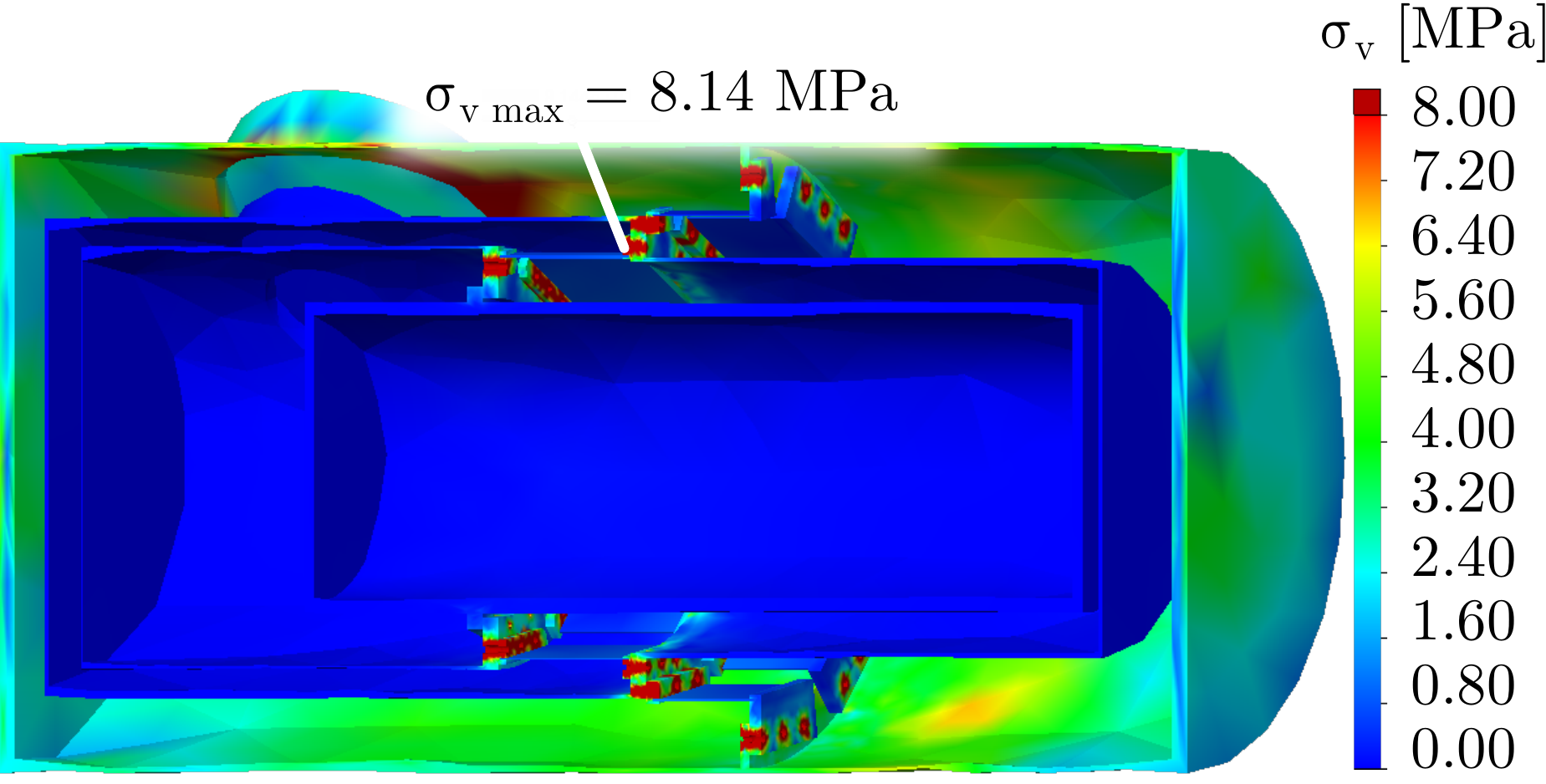}
		\caption{Von Mises Stress}
		\label{subfig:meca_simres_1}
	\end{subfigure}
	\hfill
	\begin{subfigure}[b]{0.5\textwidth}
		\centering
		\includegraphics[width=\textwidth]{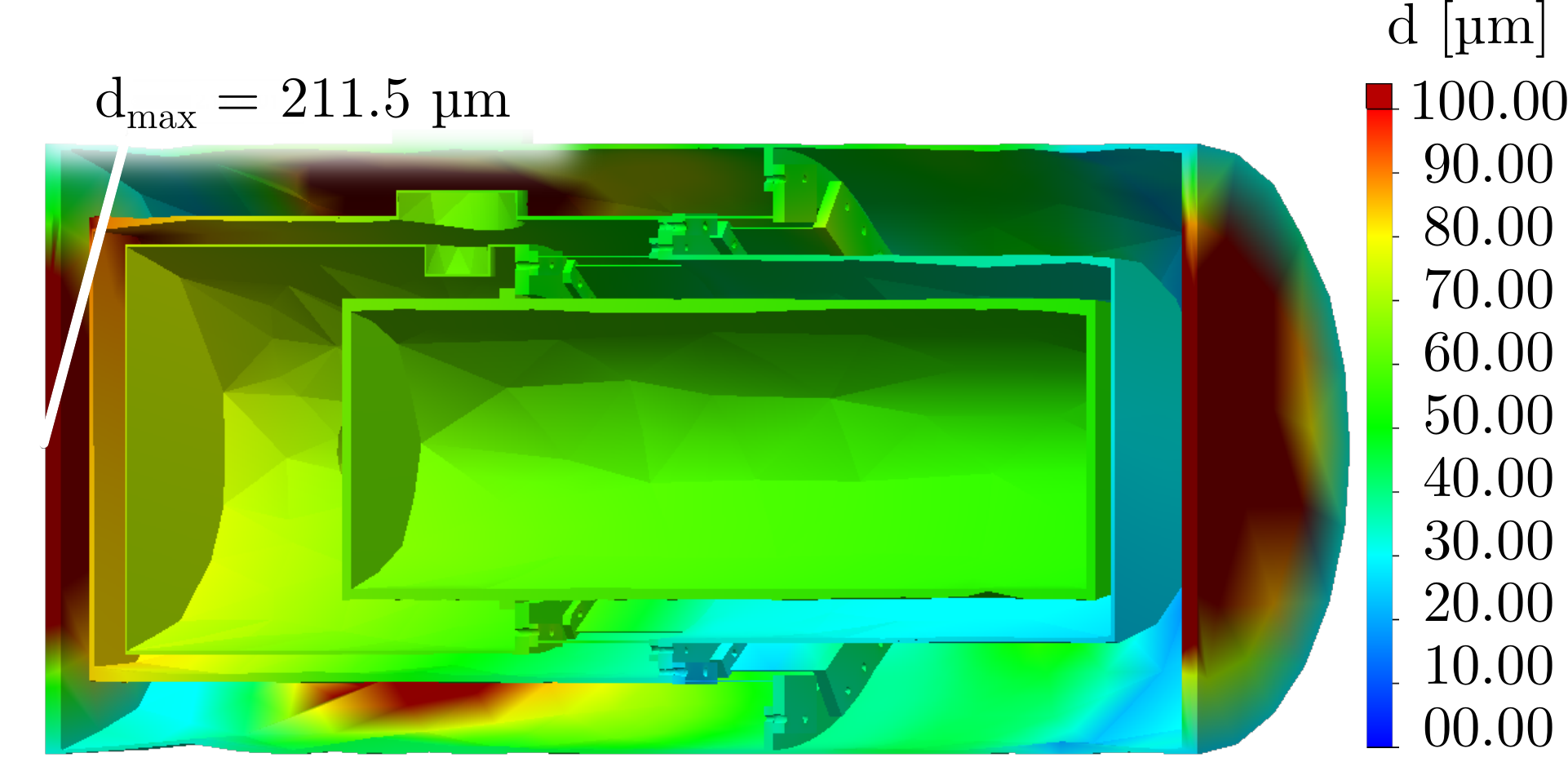}
		\caption{Displacements}
		\label{subfig:meca_simres_2}
	\end{subfigure}
	\caption{\protect\subref{subfig:meca_simres_1} Cut view of the resulting Von Mises stress applied to the CryoSim model. Maxima $\sigma_{v \: \mathrm{max}}=\SI{8.14}{\mega\pascal}$ occur at the bolted connection between the G10 tab feet and the related stage interfaces. Overall, the stress is negligible with respect to the yield strength of the test bed constituents. \protect\subref{subfig:meca_simres_2} Cut view of the resulting displacements of the CryoSim model constituents. Maxima $d_\mathrm{max}=\SI{211.5}{\micro\meter}$ are caused at the front and back of the \SIadj{300}{K} shell by the ambient pressure applied outside the cryostat and do not strongly impact the design.}
	\label{fig:meca_simres}
\end{figure}

The \gls{fea} is carried out in Solidworks using the large problem direct sparse solver, applied to the baseline CryoSim model, where each consecutive stage is supported by its predecessor via ten G10 tabs. Fig.~\ref{fig:meca_simres} shows the resulting Von Mises stress, $\sigma_v$, and displacement, $d$, for such configuration. 

The maximal stress load, $\sigma_{v \: \mathrm{max}} = \SI{8.14}{\mega\pascal}$, found at the contact surfaces of the bolts maintaining the G10 tab feet onto the rims of the connected cryostat stages, sits safely away from the yield strength of the Al 6063 and Al 6061 alloys. It appears clearly that the tabs themselves are not submitted to any strong loading in such configuration, with a maximal stress of \SI{1.84}{\mega\pascal} mostly due to buckling. To put this in perspective, tensile tests conducted for the \gls{so} \gls{sat} tabs, a design that relies solely on the G10 material for vertical support, gave a longitudinal yield strength of about \SI{200}{\mega\pascal}\cite{galitzki_simons_2024}. Tensile tests were conducted on a selection of G10 tab assemblies for Mod-Cam in Cornell University. The glue joints, made with a SCOTCH-WELD DP 2216 epoxy glue deposited with an applicator gun, did not yield as the stress was set up to \SI{3.77}{\mega\pascal}. Notwithstanding the glue connections, a design using half their number for each stage-to-stage transition will be investigated further in Section~\ref{sec:case1_g10}. Fig.~\ref{subfig:meca_simres_2}, shows the resultant of added displacements in all three directions, to scale. It illustrates that the various components are not strongly moved under load. The ambient pressured resulting from the internal vacuum results in a maximal compression of $d_\mathrm{max}=\SI{211.5}{\micro\meter}$ of the external shell while the combined weight of the stages with the G10 attachments result in tips of order \SI{30}{\micro\meter} to \SI{70}{\micro\meter}. The system has acceptable natural resonant frequencies, with the first mode occurring at \SI{71}{\giga\hertz}. The pumping line of the \gls{ptc} is the main source of mechanical vibrations, with a frequency of \SI{1.4}{\hertz}. Ambient vibrations in a laboratory are limited to a few Hertz and are quite low in amplitude. The CryoSim model displays modes away from this low set of frequencies, ensuring no resonance propagation to sensitive detectors or readout electronics. For comparison, the \gls{so} \glspl{sat} mechanical structures are designed to maintain their natural frequency above \SI{60}{\hertz} \cite{galitzki_simons_2024}.

The \gls{fea} carried on the CryoSim model, closely resembling a simplified version of Mod-Cam, presents acceptable mechanical performance to host an \gls{ot}, with safety factors of order 10 or more, displacements below \SI{100}{\micro\meter} and a first mode natural resonance occurring at \SI{71}{\giga\hertz}. As the preliminary design is validated in terms of stress and vibrations, the joint thermal analysis 
will be the subject of the next subsection. 

\subsection{Thermal analysis}
\label{subsec:thermal_gencryo}

\subsubsection{Thermal model and simulation setup} 
\label{subsubsec:thermo_setup}
\begin{figure}[h!]
    \centering
    \includegraphics[width=0.85\textwidth, keepaspectratio]{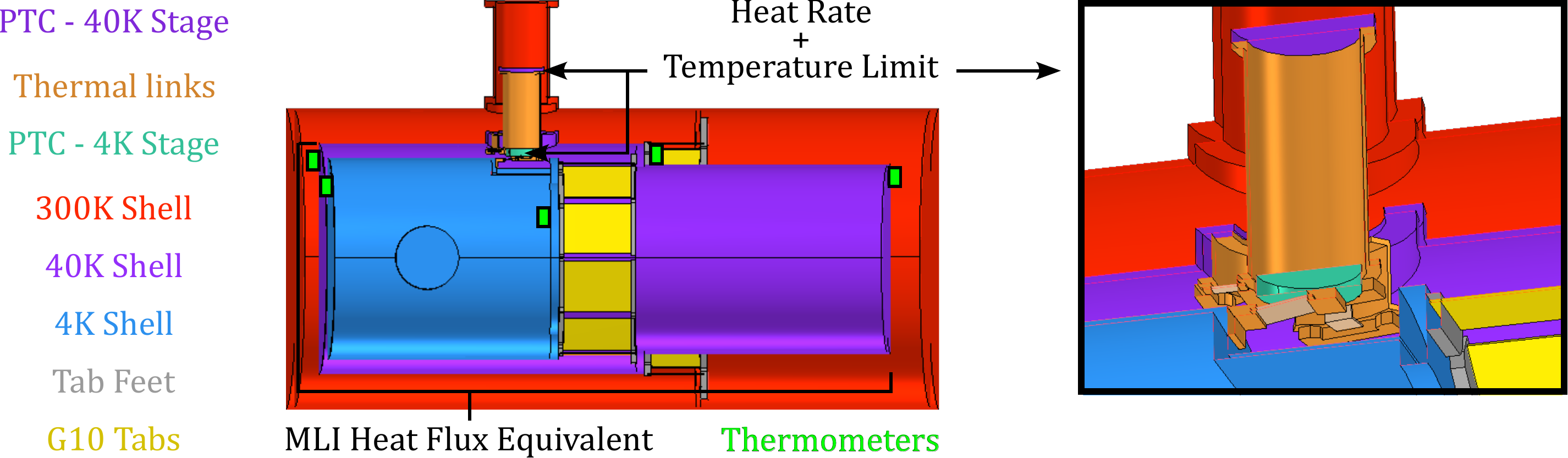}
    \caption{A cut view of CryoSim, underlining the domains defined in COMSOL Multiphysics, listed on the left, and contraptions thermally linking them, shown in the insert. Various copper blocks (green boxes) are placed in the same position as the thermometers used on Mod-Cam to monitor the simulated cooldown.}
    \label{fig:thermo_sim_setup}
\end{figure}

The CryoSim thermal model implemented in COMSOL resembles dark cooldown testing done for Mod-Cam at Cornell which did not include an \gls{ot}. As discussed in Section~\ref{subsec:overview_model} and further illustrated in Fig.~\ref{fig:thermo_sim_setup}, the dummy \gls{ot} mass is suppressed. Bolts and related holes are also removed to simplify the mesh. However, the \gls{ptc} stage plates and the thermal links connecting them to the \SIadj{40}{K} and \SIadj{4}{K} stages are now introduced to approach a realistic cooldown response, further monitored by small copper block probes positioned where thermometers are placed on Mod-Cam. 

The aluminium alloys, copper and G10 materials present non-linear thermal characteristics with temperature and frequency. As a preliminary approach, the present analysis focuses on dark testing. Namely, the \gls{ot} is removed and the cryostat windows are shut. In this context, optical loading is not considered and the cooldown and base temperatures of the cryostat can be checked. The heat conductivity, capacity and surface emissivity of each material were compiled in a \href{https://github.com/Skuggsja-Lab/skuggsja-cryobeam}{database}, shown in Figure~\ref{fig:thermprop} in Appendix~\ref{apx:sec:matprops}. The characteristics compiled were either borrowed from external sources \cite{frolec_database_2019, duthil_material_2014, nist_cryodb_2001, bock_emissivity_1995} or compiled analytically \cite{ferreira_heat_2021, king_computation_2017}. In the present context of a dark setup, the relationship between material properties and wavelengths is not incorporated to the model.


The surface-to-surface radiative coupling is performed with COMSOL Multiphysics based on a given mesh surface element view factor. It combines ambient contributions, the portion of the view from each point that is covered by the \SIadj{300}{K} ambient condition, and mutual irradiation, determined from the surrounding geometry and their local temperatures. Here the ambient contributions are not at play but would add up in a non-dark setup. The radiative coupling effectively extends the computation load drastically, especially for small features such as the copper thermal links. These links are defined as transparent bodies and the shells as opaque, meaning that no radiation is transmitted through. We believe this simplification is justified as the impact of radiative coupling is small compared to the heat flow via conduction as these components are directly fed by the \gls{ptc}.

The \gls{ptc} plates are set up with a heat rate boundary condition inputting their cooling power response \cite{coppi_cooldown_2018}. These rates are function of the temperatures $T_{40\mathrm{K}}^{\mathrm{PTC}}$ and $T_{4\mathrm{K}}^{\mathrm{PTC}}$ at the \SIadj{40}{K} and \SIadj{4}{K} plate, respectively, as depicted in Fig.~\ref{fig:ptc420} in Appendix~\ref{apxsec:ptc420}. The associated data are available in the \href{https://github.com/Skuggsja-Lab/skuggsja-cryobeam/}{CryoBeam} repository.

The \gls{mli} is set as a heat flux boundary condition on the \SIadj{40}{K} outer walls. Only the conduction contributions can be accounted for as a first order evaluation. The total flux, $q_\mathrm{MLI}$, is given by Equation~\eqref{eq:mli} \cite{ross_quantifying_2015, hedayat_analytical_2002}, where $N_{\mathrm{MLI}}$ is the number of layers, $T_\mathrm{h}$ is the high temperature seen above the \gls{mli} and $T_\mathrm{c}$ is the cold temperature at the covered surface. The heat conductivity of the \gls{mli} is given by $k_0 \kappa(T)$, where $k_0$ is a scale factor and $\kappa$ is the relative heat conductivity of the constitutive spacer material: 
\begin{equation}
    q_\mathrm{MLI} \simeq k_0 \kappa(T) \left( \frac{T_\mathrm{h} - T_\mathrm{c}}{N_\mathrm{MLI}} \right)
    \label{eq:mli}
\end{equation}
\noindent These parameters are defined accordingly to the Mod-Cam setup \cite{vavagiakis_ccatprime_2022, vavagiakis_measuring_2021}. 

For both the time domain and stationary analysis, the \gls{fea} is carried in COMSOL by a direct solver relying on a Newtonian approach modified for highly non-linear problems with the conduction heat transfer in solids and surface-to-surface radiation fully coupled. An iterative approach is used on the mesh construction, where each geometry of comparable sizes are grouped together, groups of small elements are meshed first, and the process continues further with connected groups of larger geometries. 

\subsubsection{Cooldown and stationary analysis}
\label{subsubsec:thermo_results}
The thermal analysis conducted on the CryoSim model is decomposed in two steps in the current work: a time domain study and a stationary run. The former provides a modelled cooldown response while the latter gives the stable temperature gradients. These combined results are qualitatively compared to measured data of a Mod-Cam cooldown dark as a validation of the thermal analysis of CryoSim.

The Mod-Cam cooldown temperatures were measured with cryogenic thermometers positioned at the edge of the front and back plates of each shell, and at or near the \gls{ptc} cold plates. Simulated cooldown temperatures are computed as the average temperatures, at any given time step, throughout small copper blocks positioned similarly to Mod-Cam thermometers. The model is highly non-linear as the material and \gls{mli} thermal properties and \gls{ptc} cooling powers vary as a function of temperature states. As a result, the solver does not reach convergence in the transition state to the stationary temperature gradients at the \SIadj{4}{K} stage and stops at around 40h cooldown time. The extracted temperature responses have been fitted to an inverse logistic function $T_\mathrm{l}$, described by Equation~\eqref{eq:tdiode}, where $T_\mathrm{i}$ is the temperature at the beginning of the cooldown, $T_{s}$ is the base temperature reached at the stationary stage, $\tau_\mathrm{l}$ is the time when the mid-point of the slope is reached and $k_\mathrm{l}$ is the logistic growth rate: 

\begin{equation}
    T_\mathrm{l} = \frac{T_\mathrm{i} - T_\mathrm{s}}{1 + e^{k_\mathrm{l}(t - \tau_\mathrm{l})}} +  T_\mathrm{s}
    \label{eq:tdiode}
\end{equation}

\noindent The fitted response is then extrapolated to the time when the cryostat shows a stationary response, taken from Mod-Cam cooldown data. The base temperatures $T_s$ are extracted from CryoSim stationary analysis.

\begin{figure}[h!]
	\centering
	\begin{subfigure}[b]{0.48\textwidth}
		\centering
		\includegraphics[width=\textwidth]{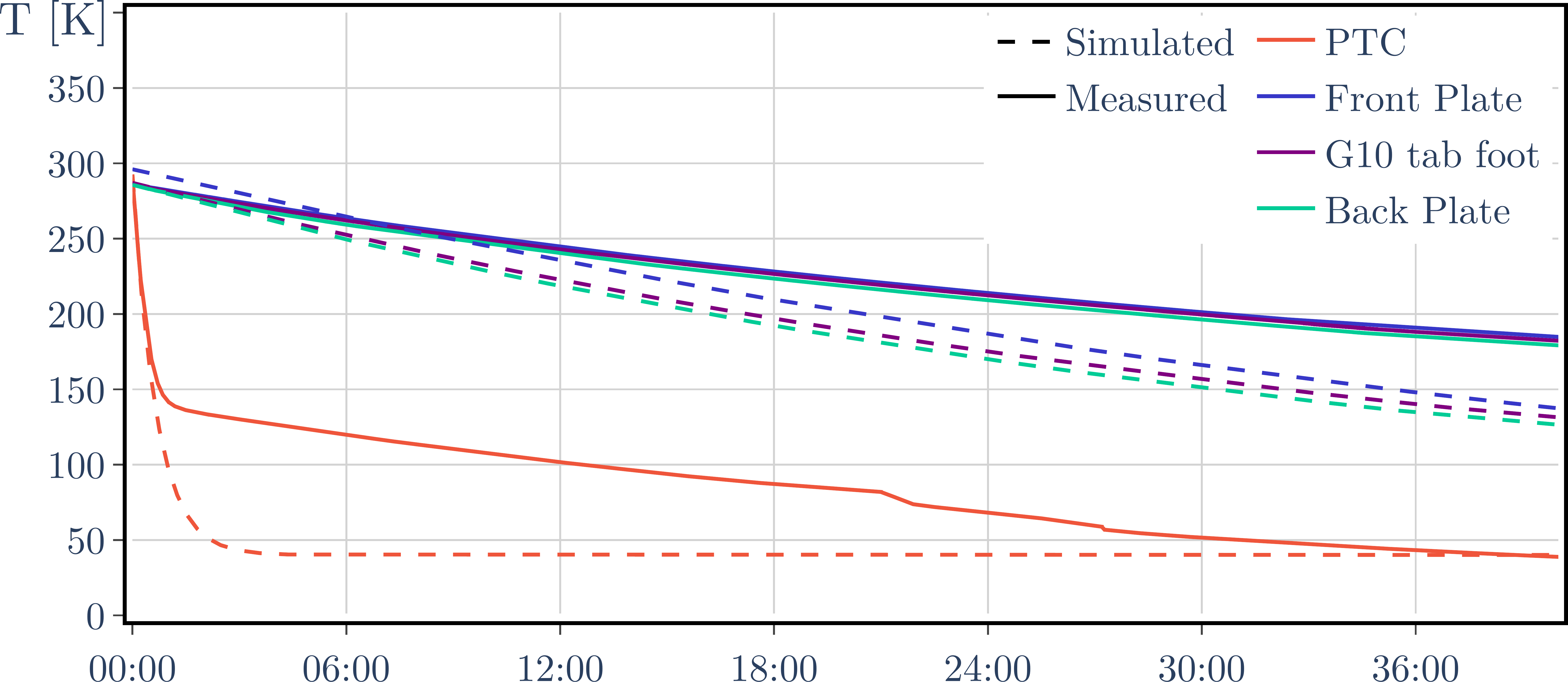}
		\caption{Temperature responses at the \SIadj{40}{K} stage}
		\label{subfig:measvssim_1}
	\end{subfigure}
        \hfill
	\begin{subfigure}[b]{0.48\textwidth}
		\centering
		\includegraphics[width=\textwidth]{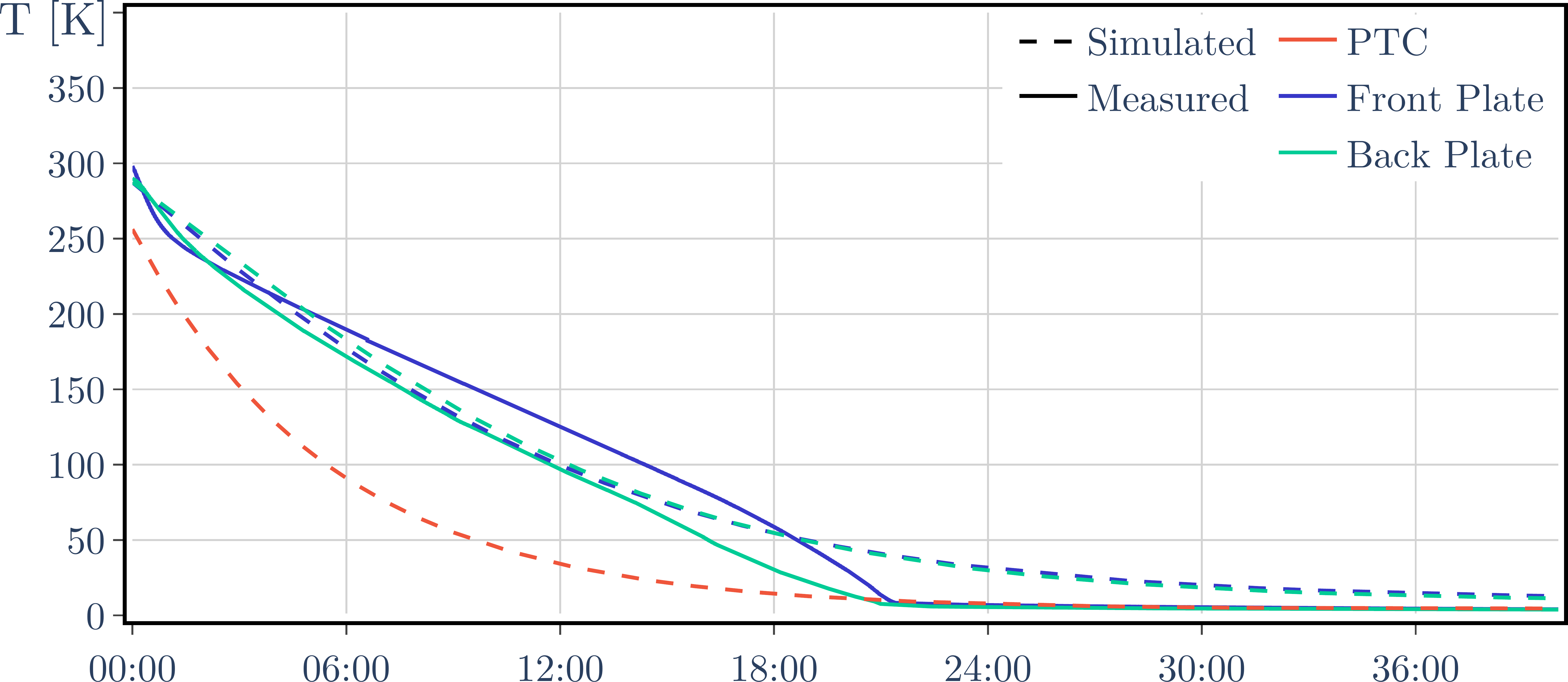}
		\caption{Temperature responses at the \SIadj{4}{K} stage}
		\label{subfig:measvssim_2}
	\end{subfigure}
	\caption{Responses of the measured (plain) and simulated (dashed) temperatures of Mod-Cam and CryoSim during a dark cooldown on \protect\subref{subfig:measvssim_1} the 
 \SIadj{40}{K} and \protect\subref{subfig:measvssim_2} the \SIadj{4}{K} stages. Whilst the model is simplified, the trend is comparable with that of the real cryostat, demonstrating the capabilities of the thermal analysis conducted.}
 \label{fig:measvssim}
\end{figure}

The modelled cooldown compares relatively well to the measurements, as is shown in Fig.~\ref{fig:measvssim}. The base temperature for the simulated and measured temperatures at the \SIadj{4}{K} stage are reached around 20h50min. For the \SIadj{40}{K} \gls{ptc} stage, the measured temperatures are taken down the thermal links whereas the simulated probe is placed at the \gls{ptc} plate. The measured and simulated temperatures for the thermometers installed on the G10 tab foot linking the \SIadj{40}{K} and the \SIadj{300}{K} stages, and the thermometers at the front and back plates also display a similar trend. The slight deviation of the slope is due to slightly different material thermal characteristics.

\begin{figure}[h!]
    \centering
    \includegraphics[width=0.5\textwidth, keepaspectratio]{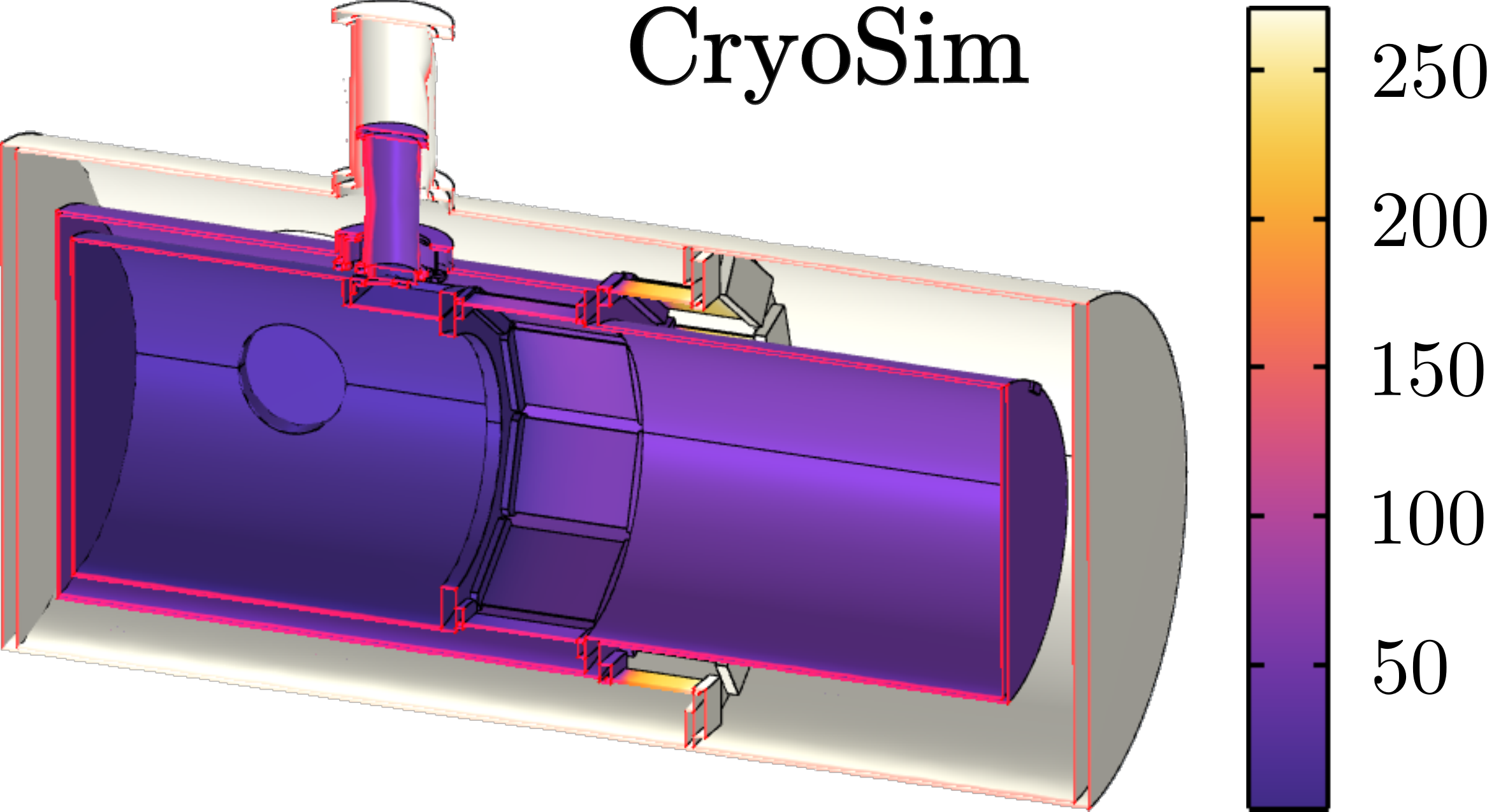}
    \caption{Temperature distribution of the full CryoSim model.}
    \label{fig:tempgrads}
\end{figure}

Figure~\ref{fig:tempgrads} shows the temperature gradients for the entire cryostat in the stationary state. Such view enables a direct assessment of critical components and potential problems that may arise during the cryostat design. For instance, it appears clearly that the G10 tabs are a direct conductive heat bridge between the connected stages.

\subsection{Discussion}
\label{subsec:model_rely}
The preliminary joint mechanical and thermal analysis conducted in this section showcases how CryoSim offers a viable tool for prototyping a cryostat capable of hosting the \glspl{ot} of upcoming \gls{cmb} experiments. The parametrisation allows one to vary the lengths, diameters, \gls{mli} lay-over or number of G10 tabs connecting two consecutive stages together, for instance, enabling the prototyping of a plethora of configurations. This approach further provides preparatory information on the stress, the natural frequencies and the thermal state of the selected system. The model presented in this work successfully compares to Mod-Cam measured data in a dark test configuration, indicating that CryoSim provides reliable stationary temperature gradients over the full cryostat under study, that can be cross-checked with a time domain analysis to asses the adequate tolerance on the convergence setup in COMSOL. To further illustrate the model capabilities, an example application will be carried in the next section, where the impact on reducing the number of G10 tabs will be discussed.

\section{CASE STUDY: LOWERING THE NUMBER OF G10 TABS}
\label{sec:case1_g10}
\begin{figure}[h!]
	\centering
	\begin{subfigure}[b]{0.49\textwidth}
		\centering
		\includegraphics[width=\textwidth]{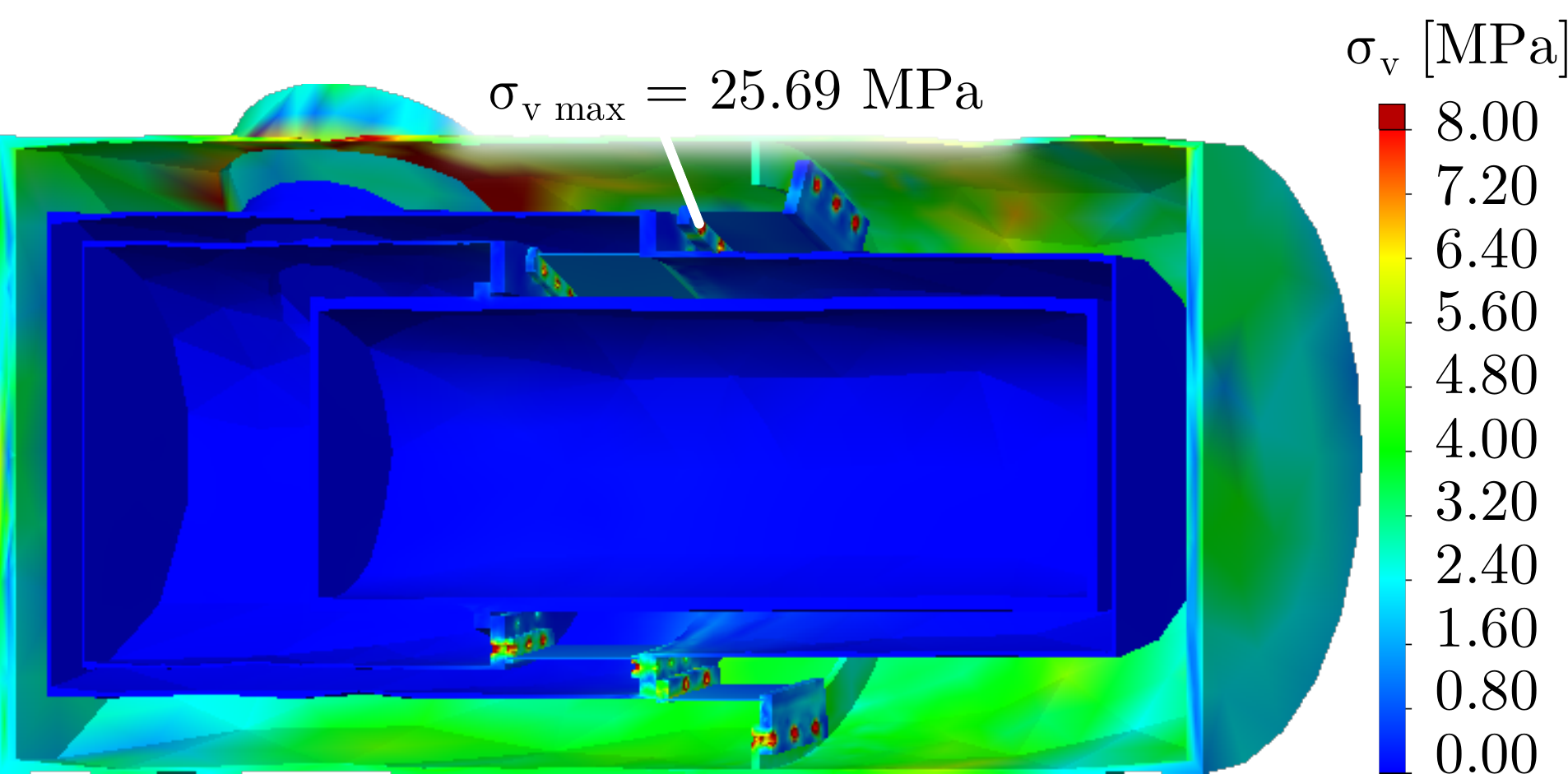}
		\caption{Von Mises Stress}
		\label{subfig:meca_5g10_1}
	\end{subfigure}
	\hfill
	\begin{subfigure}[b]{0.49\textwidth}
		\centering
		\includegraphics[width=\textwidth]{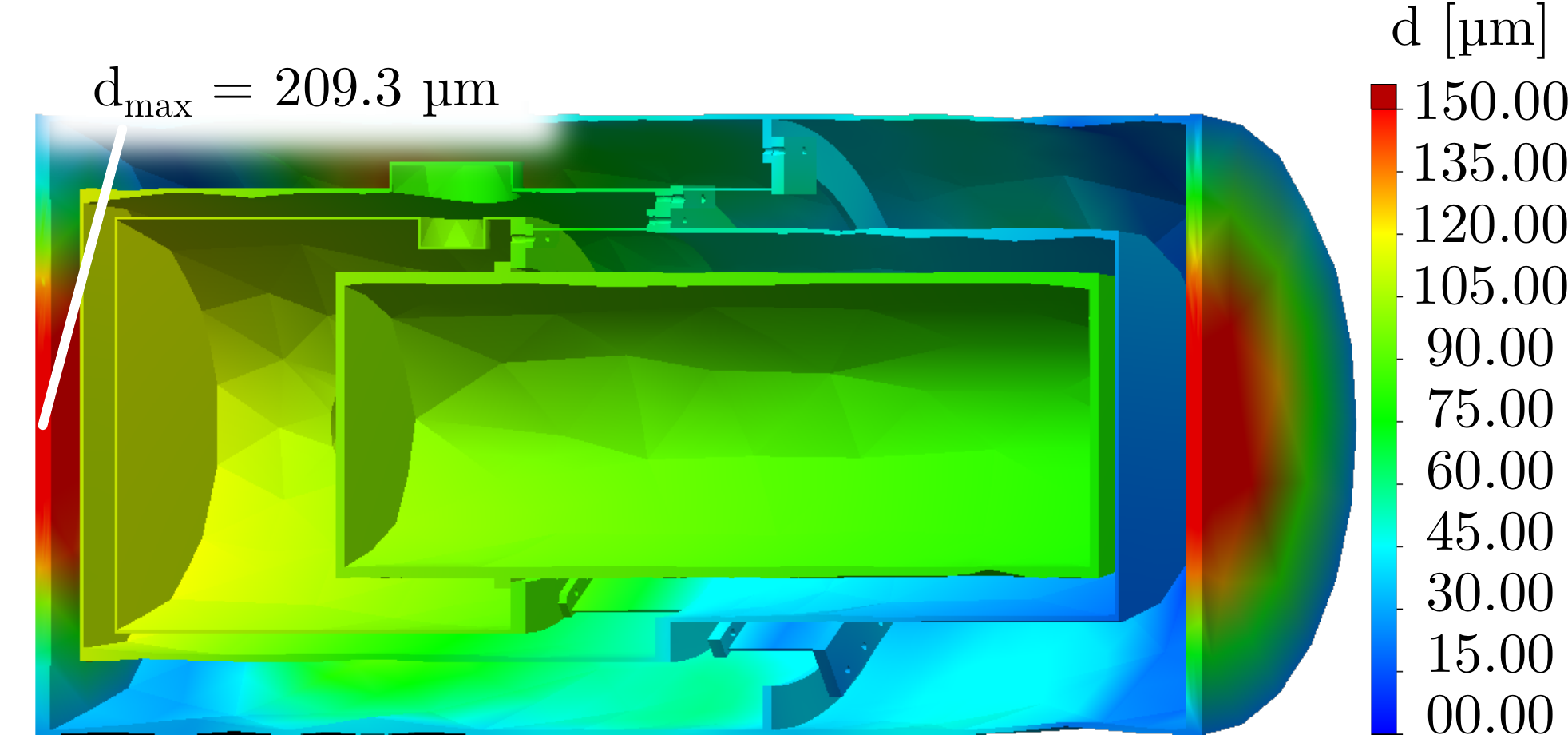}
		\caption{Displacements}
		\label{subfig:meca_5g10_2}
	\end{subfigure}
	\caption{\protect\subref{subfig:meca_5g10_1} Cut view of the resulting Von Mises stress applied to the CryoSim model. Maxima $\sigma_{v \: \mathrm{max}}=\SI{25.69}{\mega\pascal}$ occur at the bolted connection between the G10 tab feet and the related stage interfaces. \protect\subref{subfig:meca_5g10_2} Cut view of the resulting displacements of the CryoSim model constituents. Maxima $d_\mathrm{max}=\SI{209.3}{\micro\meter}$ are caused at the front and back of the \SIadj{300}{K} shell by the ambient pressure applied outside the cryostat.}
	\label{fig:meca_5g10}
\end{figure}

It was shown in Section~\ref{subsubsec:mecha_results} that ten G10 tabs is mechanically conservative. This section will use CryoSim capabilities to investigate the reduction of the number of G10 tabs by a factor of two. Figure~\ref{fig:meca_5g10} shows the static stress for CryoSim with five G10 tabs at each stage transition. 

Similarly to the ten G10 tabs case, the maximal stress load $\sigma_{v \: \mathrm{max}} = \SI{25.69}{\mega\pascal}$ occurs at the bolts contact surfaces. Whilst it is 3 times higher than the initial case study, it remains safely lower than the yield strength of the Al 6063 and Al 6061 alloys. The maximal stress resulting from buckling on the tabs themselves now sits at \SI{2.76}{\mega\pascal}, whilst measurements carried in Cornell University gave no yield of the glue joints when submitted to stresses up to \SI{3.77}{\mega\pascal}. The resulting displacements, shown to scale in Fig.~\ref{subfig:meca_5g10_2}, are doubled from what they were for ten G10 tabs, resulting in tips of order \SI{60}{\micro\meter} to \SI{120}{\micro\meter}. The first vibration mode occurs at a natural frequency of \SI{63}{\giga\hertz}.

\begin{figure}[h!]
	\centering
	\begin{subfigure}[b]{0.49\textwidth}
		\centering
		\includegraphics[width=\textwidth]{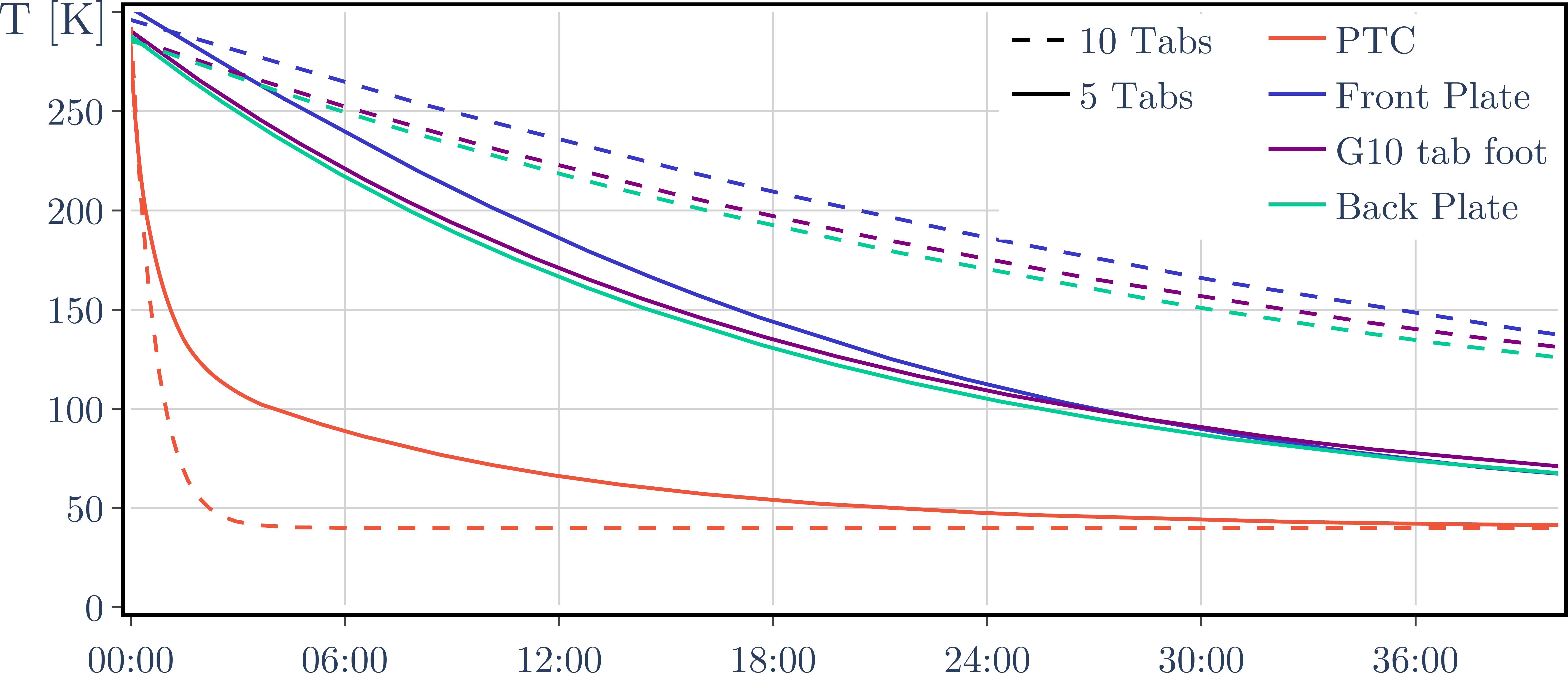}
		\caption{Simulated temperatures at the \SIadj{40}{K} stage}
		\label{subfig:5vs10sim_1}
	\end{subfigure}
 \hfill
	\begin{subfigure}[b]{0.49\textwidth}
		\centering
		\includegraphics[width=\textwidth]{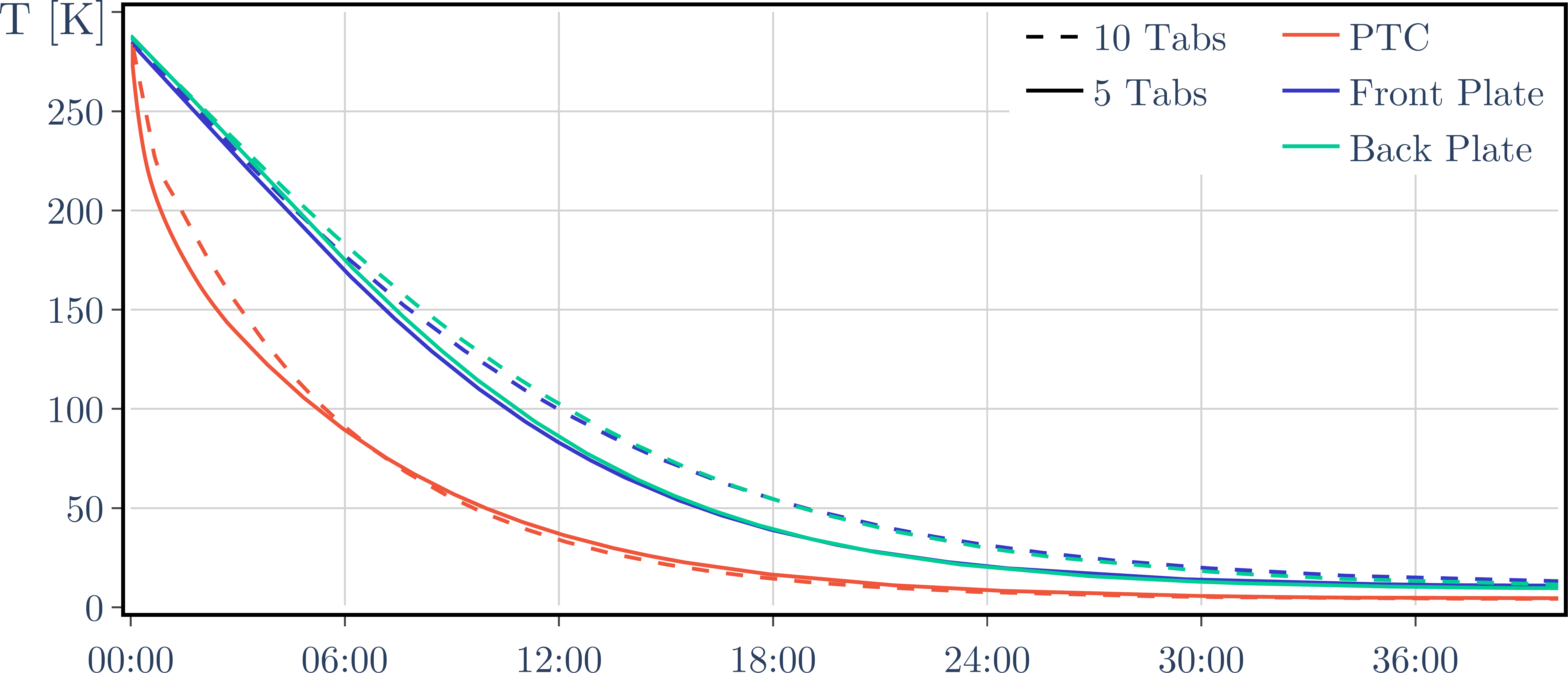}
		\caption{Simulated temperatures at the \SIadj{4}{K} stage}
		\label{subfig:5vs10sim_2}
	\end{subfigure}
	\caption{Compared responses of the simulated temperatures distributed on CryoSim for the baseline scenario with 10 tabs per stage transition against the 5 tabs case for the \protect\subref{subfig:5vs10sim_1} \SIadj{40}{K} and \protect\subref{subfig:5vs10sim_2} \SIadj{4}{K} stages.}
	\label{fig:5vs10sim}
\end{figure}

A thermal analysis of the CryoSim model with only five G10 tabs, leads to the simulated cooldown presented in Fig.~\ref{fig:5vs10sim}, showing an appreciable reduction in cooling time, nearly halved. The stationary analysis displays similar results as the ones presented in Figure~\ref{fig:tempgrads}. The joint mechanical and thermal analysis performed for this scenario would tentatively encourage a design with fewer G10 tabs, notwithstanding the glue joints, showcasing the efficiency of the method proposed.

\section{FUTURE WORK}
\label{sec:future_work}
The CryoSim model will inform the design of a test cryostat that will be assembled at the University of Iceland, which will be dedicated to the optical characterisation of \glspl{ot} for upcoming \gls{cmb} experiments. With its geometry fully parameterised, it is hoped that the CryoSim model will help the wider \gls{cmb} community in designing similar systems in an intuitive and efficient manner. Whilst facilitating a coarse preliminary cryostat design, the CryoSim model may be improved to provide more realistic thermal and mechanical characteristics, particularly by incorporating optical components into the analysis.

The mechanical studies carried in Solidworks can incorporate non-linear material responses as a function of temperature. A simplified \gls{ot}, the \gls{ptc} and the thermal links may be added to increase the accuracy of the stress and acoustics analysis, giving insights on thermal contractions for fore-optics components such as filters and lenses. The effect of the glue in the area where the G10 tabs are connected to their feet needs to be incorporated as this is likely the most fragile element. Alternative tab designs such as the one adopted for \gls{so} \glspl{sat} \cite{galitzki_simons_2024} may also be worth implementing so as to accommodate for any option preferred by the \gls{cmb} community.

Concerning the thermal analysis, incorporating the \gls{ot} is an essential evolution of the current model. It is important to note that this will increase the thermal analysis complexity as the non-linearity in material properties will extend to the frequency domain. In particular, each node of the constitutive mesh of optical components will exhibit a grey body radiative response, resulting from the absorption of incoming rays, that will then scatter within the node cone of sight. The effect of heat conduction through RF coaxial and DC lines can also be included. Incorporation of the \gls{dr} will likely be another desirable feature for the \gls{cmb} community. Implementing this further implies more profound changes to the currently simple approach taken, although provision is left on the geometry to implement such changes.

\section{CONCLUSION}
\label{sec:conclusion}
To reach the high throughput and large bandwidth of observation required to detect the faint temperature and polarisation \gls{cmb} signals, upcoming instruments designs scale up in terms of detector count, requiring larger focal plane units and more \glspl{ot} in larger cryogenic receivers.
These experiments will depend on pristine optical characterisation to ensure the polarisation purity of the measured data and the implementation of relevant correction of optical defects. Such verification needs to be carried on each individual \gls{ot} in a cold environment mimicking that of the telescope receiver.
With dedicated cryogenic test beds, such optical validation can be carried in an efficient manner. CryoSim, the set of parametrised simulation tools presented in this work, enables a straightforward preliminary design for such facilities. The reliability of the thermal model was shown in Section~\ref{sec:prelim_gencryo}, where simulated time domain temperature gradients were compared against Mod-Cam, an example of such cryogenic test bed. The CryoSim model offers a plethora of parameters one can tune, as demonstrated via a case study on the number of G10 tabs carried in Section~\ref{sec:case1_g10}. 
CryoSim may further be used to analyse potential problems that may arise with such cryostats, such as thermal leakages. Future increments of this model may incorporate a more accurate mechanical description of the materials, an \gls{ot} and a \gls{dr}, offering an efficient mean of prototyping a cryogenic test facility dedicated to the optichal characterisation of future \gls{cmb} \glspl{ot}.

\acknowledgments
Funded by the European Union (ERC, CMBeam, 101040169). Views and opinions expressed are however those of the author(s) only and do not necessarily reflect those of the European Union or the European Research Council Executive Agency. Neither the European Union nor the granting authority can be held responsible for them. \newline
Jon~E.~Gudmundsson acknowledges support from the Swedish Research Council (Reg.\ no.\ 2019-03959) and the Swedish National Space Agency (SNSA/Rymdstyrelsen). \newline
Eve~M.~Vavagiakis acknowledges support from NSF award AST-2202237.

\newpage
\appendix

\section{JOINT MECHANICAL AND THERMAL ANALYSIS - MESH PROPERTIES}
\label{apxsec:meshes}

\subsection{Mesh statistics - stress and acoustic analysis}
\label{apxsubsec:meca_meshes}
\begin{table}[h!]
	\centering
        \caption{Mesh statistics for the stress and acoustic \glspl{fea} conducted in Solidworks.}
	\resizebox{0.8\textwidth}{!}{
		\begin{tabular}{|l|c|c|}
			\hlineB{2} \hline
                 & Static Load & Acoustic \\
                \hlineB{2} \hline
                Mesh type & Blended curvature-based mesh & Mixed Mesh \\
                Max/Min element size [\si{mm}] & 215.09 / 10.75 & 216.34 / 10.82 \\
                Total nodes / elements & 105817 / 55424 & 162926 / 86350 \\
                Max aspect ratio  & 4413 & NA \\
                Elements with aspect ratio > 10 [\%] & 19.1 & NA \\
                Elements with aspect ratio < 3  [\%] & 54.5 & NA \\
                Number of \gls{dof} & 318981 & 501129 \\
			\hlineB{2} \hline
		\end{tabular}
	}
\label{tab:acou_meshprops}
\end{table}

\subsection{Mesh statistics - thermal analysis}
\label{apxsubsec:thermo_mesh}
\begin{table}[h!]
	\centering
        \caption{Mesh statistics for the thermal \gls{fea} conducted in COMSOL Multiphysics.}
	\resizebox{0.9\textwidth}{!}{
		\begin{tabular}{|l|c|c|c|c|}
                \hlineB{2} \hline
                Mesh type & Min element quality & Avg element quality & Total elements & Element volume ratio \\
                \hlineB{2} \hline
                Free Tetrahedral & \num{1.877e-3} & \num{267.4e-3} & 33021 & \num{3.492e-6} \\
			\hlineB{2} \hline
		\end{tabular}
	}
\label{tab:thermo_meshprops}
\end{table} 

\subsection{Computation capabilities - Joint mechanical and thermal analysis}
\label{apxsubsec:computing}
The system configuration used to run the joint mechanical and thermal \gls{fea} is as follows:
\begin{itemize}
    \item[] \textbf{OS} - Microsoft Windows 11 Education Version 10.0.22631 Build 22631
    \item[] \textbf{Processor} - 12th Gen Intel(R) Core(TM) i9-12900K, 3200 Mhz, 16 Core(s), 24 Logical Processor(s)
    \item[] \textbf{Physical Memory} - 31.8 GB
    \item[] \textbf{Virtual Memory} - 81.8 GB
    \item[] \textbf{GPU Resources} - NVIDIA Quadro P400 \& Intel(R) PEG10 - 460D
\end{itemize}
Considering the meshes depicted in Appendix~\ref{apxsec:meshes}, the subsequent performance was reached:
\begin{itemize}
    \item[] \textbf{Mechanical \gls{fea}} - Static analysis run time: 00:07:25; Frequency analysis run time: 00:00:43
    \item[] \textbf{Thermal \gls{fea}} - Time Domain for $t \in [\text{00:00:00};\: \text{21:40:00}]$ with \SI{10}{min} increments and a tolerance factor for convergence set at 0.01 - Run time: 00:26:20; Stationary with a tolerance factor for convergence set at 0.01: 00:35:18.
\end{itemize}

\clearpage
\section{MATERIAL PROPERTIES DATABASE}
\label{apx:sec:matprops}
\begin{figure}[h!]
	\centering
	\begin{subfigure}[b]{0.45\textwidth}
		\centering
		\includegraphics[width=\textwidth]{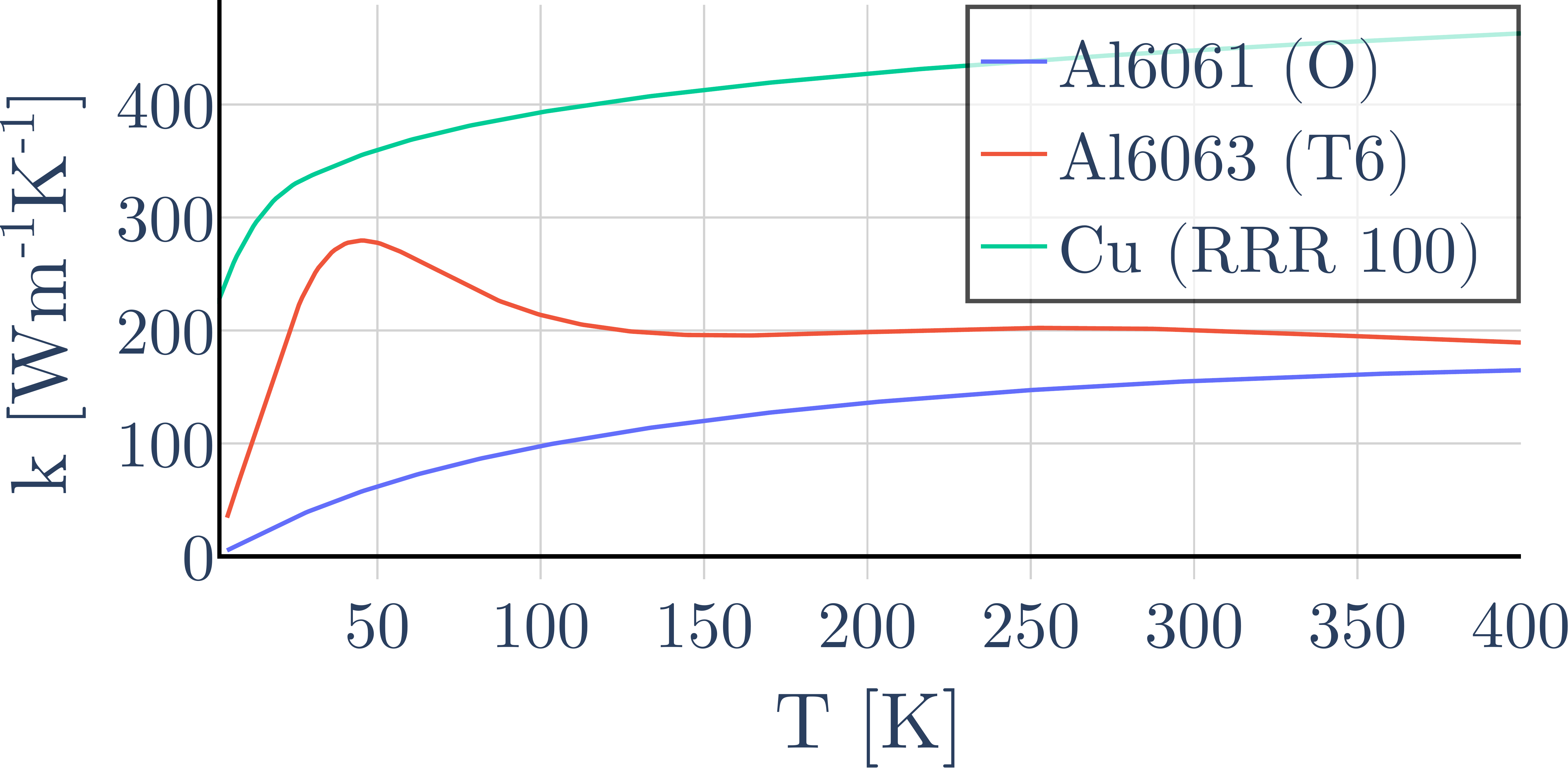}
		\caption{Heat conductivity $k$ of Al 6063, Al 6061 and low purity copper.}
		\label{subfig:thermprop_1}
	\end{subfigure}
	\hfill
	\begin{subfigure}[b]{0.45\textwidth}
		\centering
		\includegraphics[width=\textwidth]{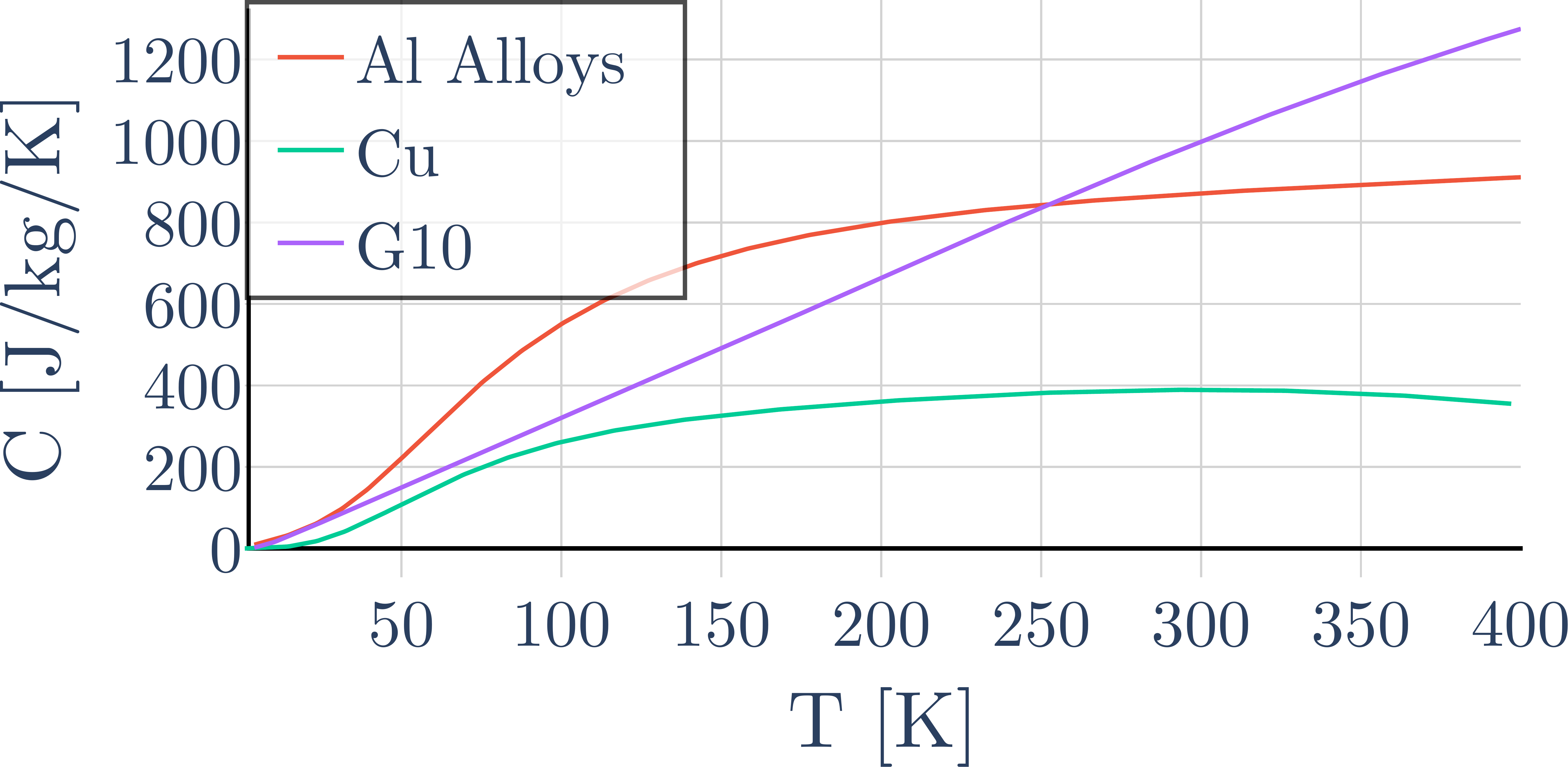}
		\caption{Heat capacity $C$ of Aluminium alloys, copper and G10.}
		\label{subfig:thermprop_2}
	\end{subfigure}
        \\
	\begin{subfigure}[b]{0.45\textwidth}
		\centering
		\includegraphics[width=\textwidth]{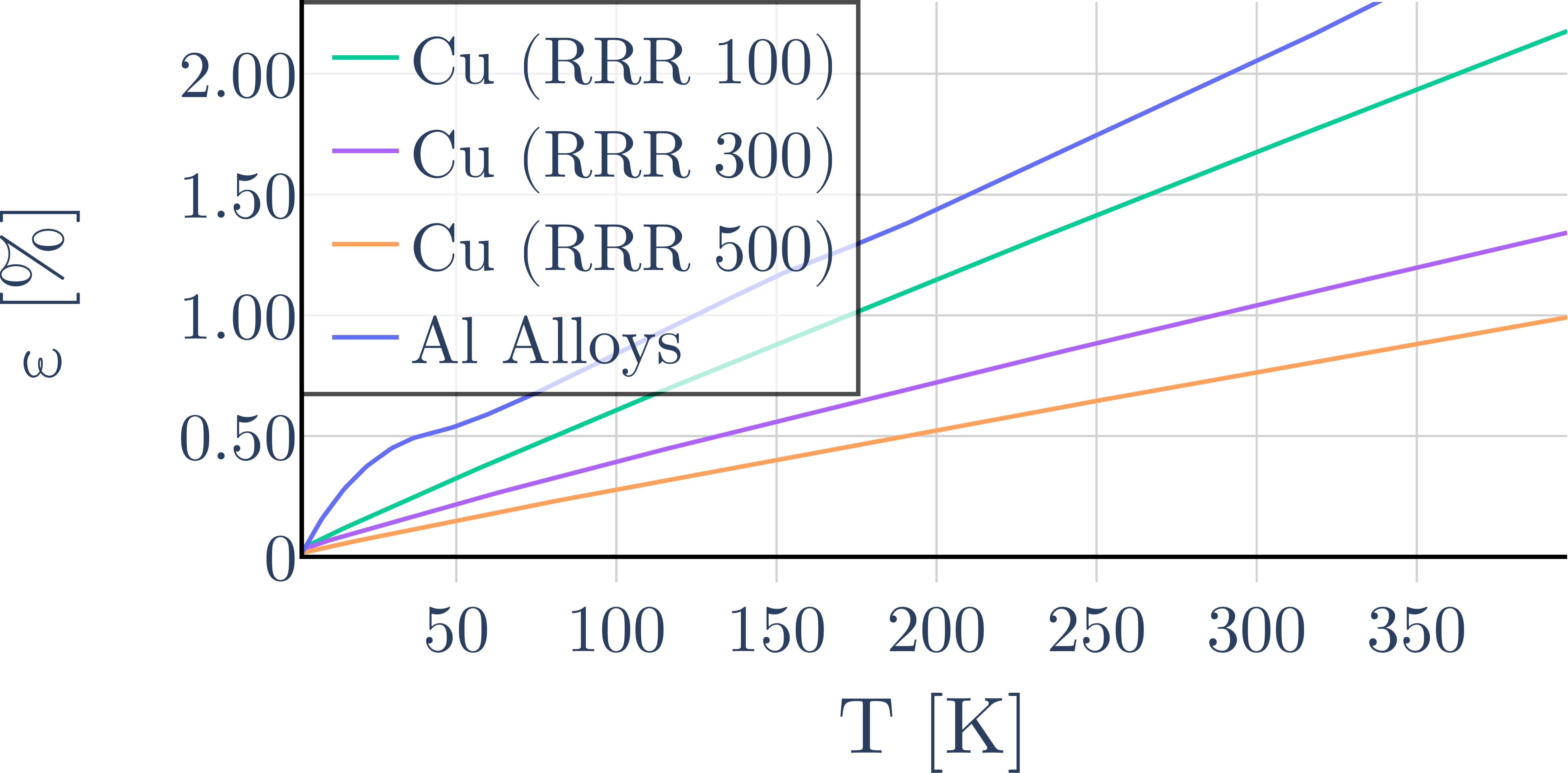}
		\caption{Surface emissivity $\varepsilon$ of copper of various purity and aluminium alloys.}
		\label{subfig:thermprop_3}
	\end{subfigure}
	\caption{\protect\subref{subfig:thermprop_1} Heat conductivity, \protect\subref{subfig:thermprop_2} heat capacity and \protect\subref{subfig:thermprop_3} surface emissivity of a sample of materials used for the CryoSim model, showcasing their thermal non-linearity.}
	\label{fig:thermprop}
\end{figure}

\section{POWER SURFACES OF THE PTC 420}
\label{apxsec:ptc420}
\begin{figure}[h!]
	\centering
	\begin{subfigure}[b]{0.4\textwidth}
		\centering
		\includegraphics[width=\textwidth]{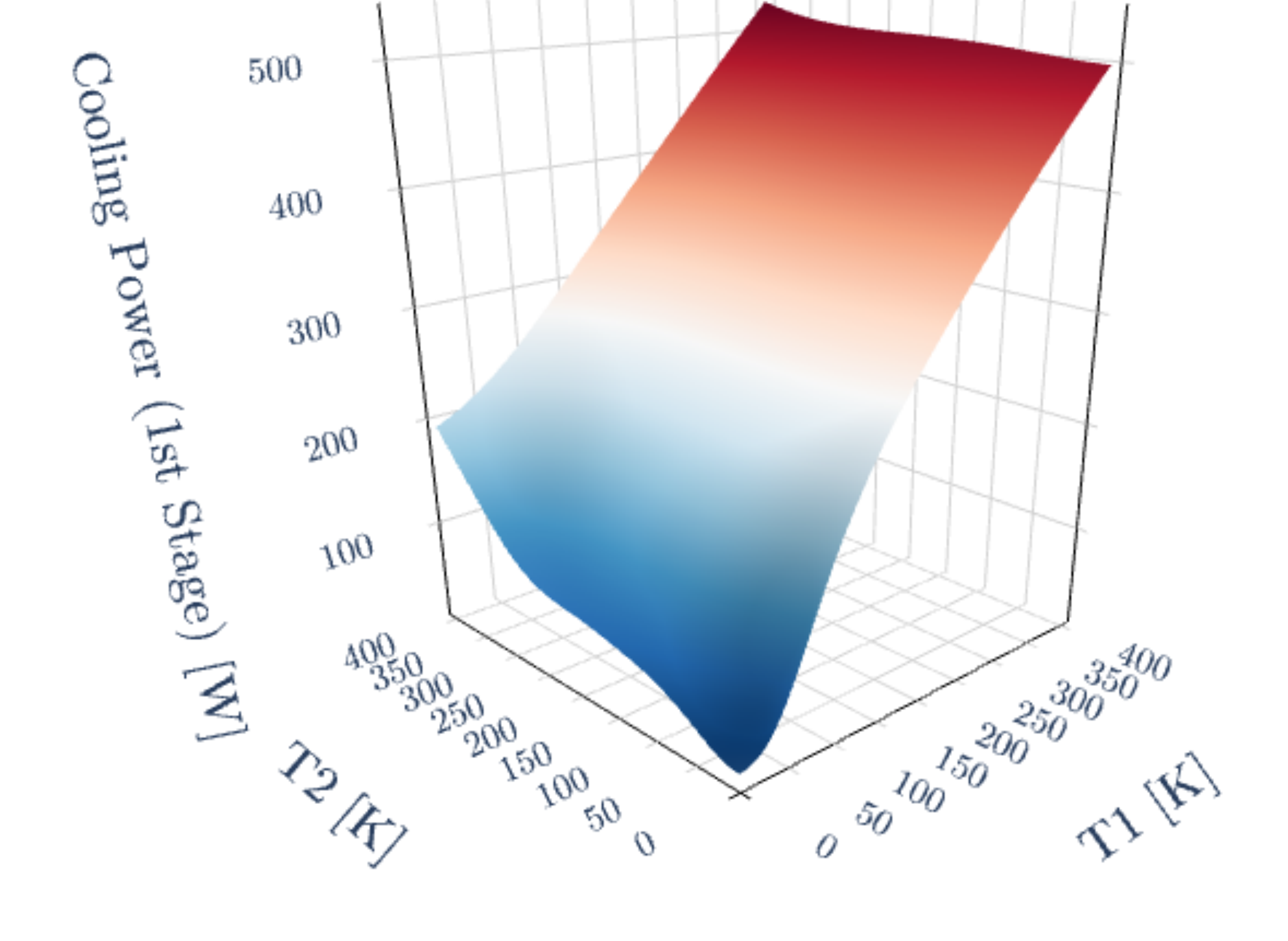}
		\caption{Cooling Power of the 1\textsuperscript{st} stage}
		\label{subfig:ptc420s1}
	\end{subfigure}
	\hfill
	\begin{subfigure}[b]{0.5\textwidth}
		\centering
		\includegraphics[width=\textwidth]{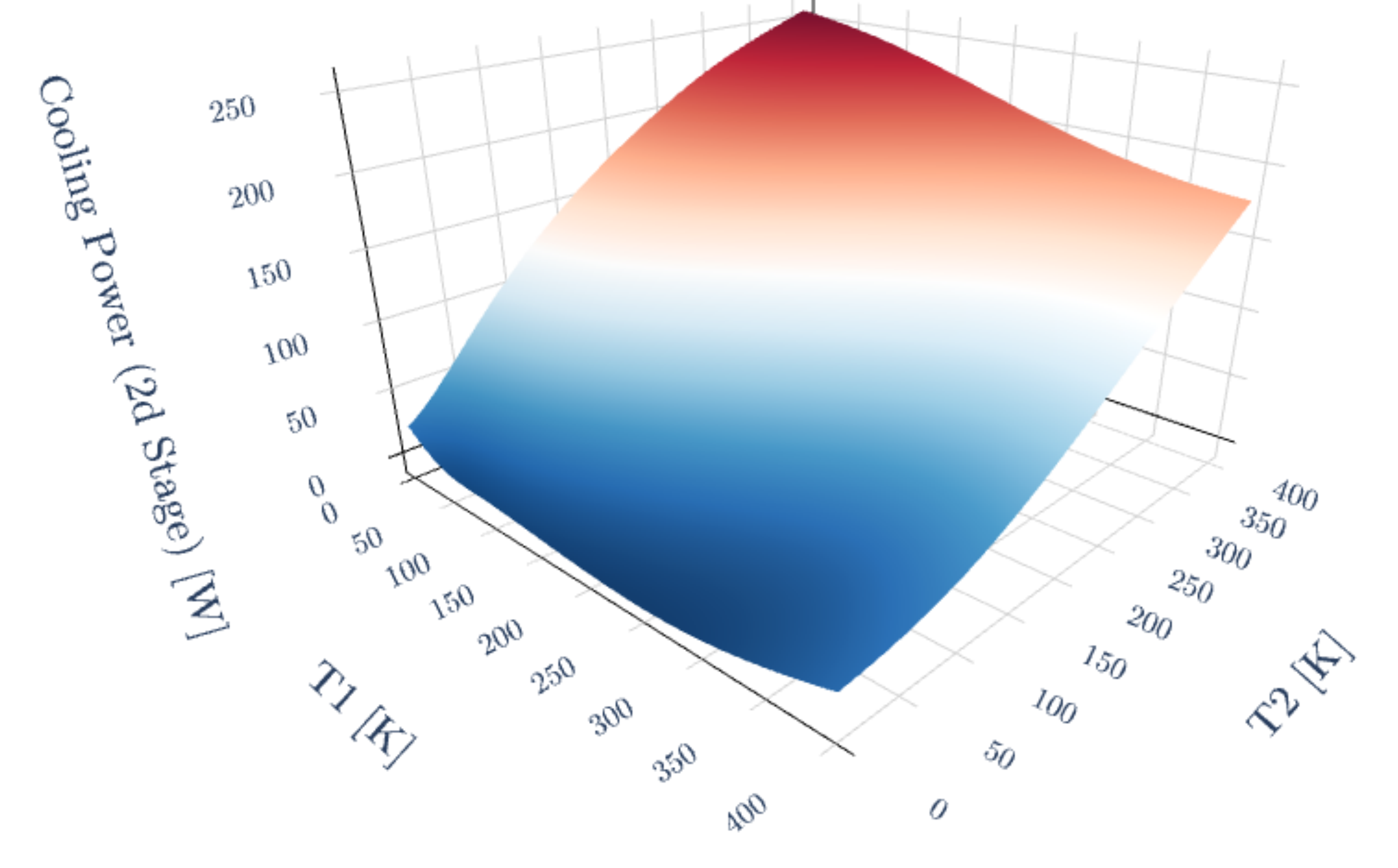}
		\caption{Cooling Power of the 2\textsuperscript{d} stage}
		\label{subfig:ptc420s2}
	\end{subfigure}
	\caption{Cooling power surfaces of the \protect\subref{subfig:ptc420s1} 1\textsuperscript{st} and \protect\subref{subfig:ptc420s2} 2\textsuperscript{d} stages of the Bluefors-Cryomech \gls{ptc} 420, function of the temperature of each stage.}
	\label{fig:ptc420}
\end{figure}

\clearpage
\bibliography{bibliography} 
\bibliographystyle{spiebib} 

\end{document}